\documentclass[aps]{revtex4}

\usepackage{amsmath}
\usepackage{stmaryrd}

\def\ov#1{\overline{#1}}

\def\wt#1{\widetilde{#1}}
\def\vb#1{\mbox{\boldmath$#1$}}
\def\pd#1#2{\frac{\partial #1}{\partial #2}}
\def\fd#1#2{\frac{\delta #1}{\delta #2}}
\def\wh#1{\widehat{#1}}
\def\bdot{\,\vb{\cdot}\,}
\def\btimes{\,\vb{\times}\,}

\def\bhat{\wh{{\sf b}}}
\def\cal#1{\mathcal{#1}}

\def\bhat{\wh{{\sf b}}}
\def\exd{{\sf d}}

\newcommand{\bc}{\begin{center}}
\newcommand{\ec}{\end{center}}
\newcommand{\bt}{\begin{tabbing}}
\newcommand{\et}{\end{tabbing}} 
\newcommand{\be}{\begin{eqnarray*}}
\newcommand{\ee}{\end{eqnarray*}}
\newcommand{\bs}{\begin{slide}}
\newcommand{\es}{\end{slide}}

\begin{document}

\title{Exact momentum conservation laws for the gyrokinetic Vlasov-Poisson equations}

\author{Alain J.~Brizard$^{1}$ and Natalia Tronko$^{2,3}$}
\affiliation{$^{1}$Department of Chemistry and Physics, Saint Michael's College, Colchester, VT 05439, USA \\ $^{2}$Centre de Physique Th\'{e}orique, 
Campus de Luminy, Case 907, 13288 Marseille cedex 9, France \\ $^{3}$Centre for Fusion, Space and Astrophysics, Department of Physics, University of Warwick, Coventry CV4 7AL, UK} 

\begin{abstract}
The exact momentum conservation laws for the nonlinear gyrokinetic Vlasov-Poisson equations are derived by applying the Noether method on the gyrokinetic variational principle [A.~J.~Brizard, Phys.~Plasmas {\bf 7}, 4816 (2000)]. From the gyrokinetic Noether canonical-momentum equation derived by the Noether method, the gyrokinetic parallel momentum equation and other gyrokinetic Vlasov-moment equations are obtained. In addition, an exact gyrokinetic toroidal angular-momentum conservation law is derived in axisymmetric tokamak geometry, where the transport of parallel-toroidal momentum is related to the radial gyrocenter polarization, which includes contributions from the guiding-center and gyrocenter transformations. 
\end{abstract}

\begin{flushright}
July 10, 2011
\end{flushright}


\maketitle

\section{Introduction}

The property of exact energy conservation has long been a stringent test on the accuracy of nonlinear gyrokinetic numerical simulation codes 
\cite{Garbet_NF}. While the gyrokinetic energy conservation law was initially constructed directly from the nonlinear gyrokinetic equations 
\cite{Dubin_1983,HLB_1988,Brizard_1989a,Brizard_1989b}, a more general approach relies on the existence of a variational formulation for the nonlinear gyrokinetic equations \cite{Sugama,Brizard_VP,Brizard_gyVP,Brizard_Hahm,Brizard_2010} and the use of the Noether method \cite{Brizard_JPP} associated with the symmetry of the gyrokinetic Lagrangian density with respect to infinitesimal time translations.

The turbulent transport equations for parallel momentum and toroidal angular momentum in magnetized plasmas have recently been the focus of intense investigation \cite{PHD_2009}. These gyrokinetic momentum transport equations are generally derived as moment equations of the nonlinear gyrokinetic Vlasov equation \cite{TSH_2007,McD_2009_PoP,McD_2009_PRL,Parra_Catto,Scott_Smirnov}. The present paper is concerned with an alternative derivation of the momentum and angular momentum conservation laws for the nonlinear gyrokinetic Vlasov-Poisson equations by the Noether method, which associates a conservation law to each symmetry of the gyrokinetic Lagrangian density with respect to infinitesimal space-time translations and rotations. The energy-momentum conservation laws associated with the guiding-center Vlasov-Maxwell equations have previously been investigated with a variational formulation by Pfirsch and Morrison \cite{PM_85}. The present work generalizes Pfirsch and Morrison's work \cite{PM_85} to the gyrokinetic Vlasov-Poisson equations, first, and the gyrokinetic Vlasov-Maxwell equations in a future paper.

The gyrokinetic Noether canonical-momentum equation derived for the gyrokinetic Vlasov-Poisson equations is an exact equation that describes the transport of gyrokinetic canonical momentum associated with the gyrocenter particles (represented by the gyrocenter Vlasov distribution) and the electrostatic fluctuations (represented by the gyrokinetic Poisson equation), in the presence of a nonuniform background magnetic field. While momentum can be exchanged between the gyrokinetic Vlasov-Poisson fields and the background magnetic field when the latter does not possess exact spatial symmetries (e.g., toroidal and poloidal directions in stellarator geometry \cite{PH_1,PH_2,PH_3}), the canonical toroidal angular momentum carried by the gyrokinetic Vlasov-Poisson fields is conserved exactly in an axisymmetric tokamak magnetic field, which allows the transfer of momentum between the gyrocenter Vlasov distribution and the electrostatic-field fluctuations. From the gyrokinetic Noether canonical-momentum equation derived by the Noether method, the gyrokinetic parallel momentum equation and other gyrokinetic Vlasov-moment equations are obtained. 

The remainder of the paper is organized as follows. In Sec.~\ref{sec:gvp}, we review the variational derivations \cite{Brizard_gyVP,Brizard_Hahm,Brizard_2010} of the nonlinear gyrocenter canonical and noncanonical Hamilton equations and the nonlinear gyrokinetic Vlasov-Poisson equations. In Sec.~\ref{sec:gyro_mom}, we apply the Noether method (briefly reviewed in App.~\ref{sec:Noether}) to derive the exact gyrokinetic linear and angular momentum conservation laws. In Sec.~\ref{sec:apply}, we present an application of the gyrokinetic angular momentum conservation law by deriving the parallel-toroidal gyrokinetic momentum conservation law in axisymmetric tokamak geometry. Lastly, we summarize our work and discuss future work in Sec.~\ref{sec:summ}.

\section{\label{sec:gvp}Gyrokinetic Variational Principles}

The variational derivations of the nonlinear gyrokinetic Vlasov-Poisson and Vlasov-Maxwell equations and the gyrocenter Hamilton equations guarantee that these equations possess important conservation laws such as energy and momentum. The purpose of this Section is to review these variational principles before they are used in the remainder of the paper. 

\subsection{Gyrocenter Hamilton variational principle}

The single-particle variational principle used for the derivation of the gyrocenter Hamiltonian dynamics of a charged particle of mass $m$ and charge $e$ is based on the noncanonical phase-space Lagrangian \cite{Brizard_1989a,Brizard_Hahm}
\begin{equation}
\Gamma_{\rm gy} \;=\; \left( \frac{e}{c}\,{\bf A} \;+\; p_{\|}\,\bhat \right)\bdot\exd{\bf X} \;+\; (mc/e)\,\mu\,\exd\zeta \;-\; H_{\rm gy}\;\exd t
\;\equiv\; \Gamma_{\alpha}\;\exd Z^{\alpha} \;-\; H_{\rm gy}\;\exd t,
\label{eq:Gamma_gy}
\end{equation}
where the background magnetic field ${\bf B} \equiv \nabla\btimes{\bf A}$ is defined in terms of the time-independent magnetic vector potential 
${\bf A}$, while the gyrocenter phase-space coordinates are the gyrocenter position ${\bf X}$, the gyrocenter parallel momentum $p_{\|}$, the gyrocenter magnetic moment $\mu$, and the gyrocenter gyroangle $\zeta$. The gyrocenter Hamiltonian $H_{\rm gy}$ is defined as
\begin{equation}
H_{\rm gy}({\bf X}, p_{\|}, \mu, t;\; \phi_{1}) \;\equiv\; H_{\rm gc}({\bf X}, p_{\|}, \mu) + e\,\Phi_{\rm gy}({\bf X},\mu, t),
\label{eq:H_gy}
\end{equation}
where $H_{\rm gc} \equiv \mu\,B + p_{\|}^{2}/2m$ denotes the (unperturbed) guiding-center Hamiltonian. The gyrocenter potential $\Phi_{\rm gy}$ is defined in terms of the first-order fluctuating guiding-center electrostatic potential
\begin{equation}
\phi_{1{\rm gc}} \;\equiv\; {\sf T}_{\rm gc}^{-1}\phi_{1} \;=\; \phi_{1}({\bf X} + \vb{\rho}_{\rm gc},t), 
\label{eq:Phi1_gc_def}
\end{equation}
as
\begin{equation}
\Phi_{\rm gy} \;\equiv\; \epsilon\,\langle\phi_{1{\rm gc}}\rangle \;-\; \frac{\epsilon^{2}}{2}\;\left\langle \left\{ S_{1},\frac{}{}\phi_{1{\rm gc}}\right\}_{\rm gc}\right\rangle \;+\; \cdots,
\label{eq:Psi_gy_def}
\end{equation}
where $\vb{\rho}_{\rm gc}$ denotes the guiding-center gyroradius (which contains higher-order contributions due to the nonuniformity of the background magnetic field \cite{RGL,Cary_Briz}), $\langle\cdots\rangle$ denotes the gyroangle-averaging operation, and the guiding-center Poisson bracket $\{\;,\;\}_{\rm gc}$ is defined as \cite{RGL,Cary_Briz}
\begin{eqnarray}
\{F,\; G\}_{\rm gc} & = & \frac{\Omega}{B}\left(\pd{F}{\zeta}\,\pd{G}{\mu} - \pd{F}{\mu}\,\pd{G}{\zeta}\right) \;+\; \frac{{\bf B}^{*}}{B_{\|}^{*}}
\bdot\left( \nabla F\,\pd{G}{p_{\|}} - \pd{F}{p_{\|}}\,\nabla G \right) \;-\; \frac{c\bhat}{eB_{\|}^{*}}\bdot\nabla F\btimes\nabla G,
\label{eq:PB_gc}
\end{eqnarray} 
where 
\begin{equation}
{\bf B}^{*} \;\equiv\; \nabla\btimes{\bf A}^{*} \;=\; {\bf B} + (cp_{\|}/e)\,\nabla\btimes\bhat \;=\; B_{\|}^{*}\;\bhat \;+\; p_{\|}\;\frac{c\bhat}{e}\btimes\left(\bhat\bdot\nabla\bhat\right)
\label{eq:Bstar_def}
\end{equation}
and $B_{\|}^{*} \equiv \bhat\bdot{\bf B}^{*} = B + (cp_{\|}/e)\;\bhat\bdot\nabla\btimes\bhat$. The second-order ponderomotive potential in 
Eq.~\eqref{eq:Psi_gy_def} is expressed in terms of the first-order function $S_{1}$, which is formally defined as the solution of the first-order equation 
\cite{Dubin_1983,Brizard_Hahm}
\begin{equation}
\frac{d_{\rm gc}S_{1}}{dt} \;=\; e\,\wt{\phi}_{1{\rm gc}} \;\equiv\; e\, \left( \phi_{1{\rm gc}} \;-\frac{}{} \langle\phi_{1{\rm gc}}\rangle \right),
\label{eq:S1_ODE}
\end{equation}
where $d_{\rm gc}/dt \equiv \partial/\partial t + \{\;, H_{\rm gc}\}_{\rm gc}$ denotes the unperturbed (guiding-center) Vlasov operator. To lowest order in the gyrokinetic space-time-scale ordering \cite{Brizard_Hahm}, the solution for $S_{1}$ is approximated as
\begin{equation}
S_{1} \;\simeq\; \frac{e}{\Omega}\;\int\,\wt{\phi}_{1{\rm gc}}\;d\zeta,
\label{eq:S1_def}
\end{equation}
where higher-order terms (not considered in standard gyrokinetic theory \cite{Brizard_Hahm}) have been derived in Ref.~\cite{Brizard_Mishchenko}. We note that, while the gyroangle-independent potential $\langle\phi_{1{\rm gc}}\rangle$ contributes to the linear (first-order) perturbed gyrocenter Hamiltonian dynamics, the gyroangle-dependent potential $\wt{\phi}_{1{\rm gc}}$ contributes to the (second-order) gyrocenter ponderomotive Hamiltonian in Eq.~\eqref{eq:H_gy} as well as gyrocenter polarization effects [see Eq.~\eqref{eq:rho1_gy} below]. Lastly, high-order terms in Eq.~\eqref{eq:Psi_gy_def} have recently been derived at third order in Ref.~\cite{Mish_Brizard}, which yield nonlinear gyrocenter polarization effects.

\subsubsection{Gyrocenter noncanonical Euler-Lagrange equations}

The noncanonical gyrocenter equations of motion are obtained from the variational principle 
\begin{equation}
0 \;=\; \delta\int\Gamma_{\rm gy} \;\equiv\; \int\;\delta Z^{\alpha}\;\left(\omega_{\alpha\beta}\;\frac{d_{\rm gy}Z^{\beta}}{dt} \;-\; 
\pd{H_{\rm gy}}{Z^{\alpha}} \right)\;dt,
\label{eq:gy_EL_VP}
\end{equation}
which, for arbitrary virtual displacements $\delta Z^{\alpha}$ in gyrocenter phase space (that vanish at the integration boundaries), yields the noncanonical gyrocenter Euler-Lagrange equations
\begin{equation}
\omega_{\alpha\beta}\;\frac{d_{\rm gy}Z^{\beta}}{dt} \;=\; \pd{H_{\rm gy}}{Z^{\alpha}},
\label{eq:gy_EL}
\end{equation}
where the antisymmetric Lagrange matrix $\omega_{\alpha\beta} \equiv \partial_{\alpha}\Gamma_{\beta} - \partial_{\beta}\Gamma_{\alpha}$ is derived from the symplectic part of the phase-space Lagrangian \eqref{eq:Gamma_gy}, while the gyrocenter Vlasov operator $d_{\rm gy}/dt \equiv \partial/\partial t + \{\;, H_{\rm gy}\}_{\rm gc}$ is defined in terms of the gyrocenter Hamiltonian \eqref{eq:H_gy} and the guiding-center Poisson bracket \eqref{eq:PB_gc}. The noncanonical Euler-Lagrange equations \eqref{eq:gy_EL} for the reduced gyrocenter coordinates $({\bf X}, p_{\|})$  are
\begin{eqnarray}
\frac{e}{c}\;\frac{d_{\rm gy}{\bf X}}{dt}\btimes{\bf B}^{*} \;-\; \frac{d_{\rm gy}p_{\|}}{dt}\;\bhat & = & \nabla H_{\rm gy}, 
\label{eq:EL_X}
\end{eqnarray}
and
\begin{eqnarray}
\bhat\bdot\frac{d_{\rm gy}{\bf X}}{dt} & = & \pd{H_{\rm gy}}{p_{\|}} \;=\; \frac{p_{\|}}{m},
\label{eq:EL_p}
\end{eqnarray}
where the gradient operator in Eq.~\eqref{eq:EL_X} is evaluated at constant $p_{\|}$ and $\mu$.

\subsubsection{Gyrocenter noncanonical Hamilton equations}

By inverting the Lagrange matrix $\omega_{\alpha\beta}$ in Eq.~\eqref{eq:gy_EL}, we obtain the guiding-center Poisson matrix $J^{\alpha\beta} \equiv \{ 
Z^{\alpha},\; Z^{\beta}\}_{\rm gc}$ from which Eq.~\eqref{eq:PB_gc} is constructed. Through this inversion, the gyrocenter Euler-Lagrange equations 
\eqref{eq:gy_EL} are converted into the gyrocenter Hamilton equations
\begin{equation}
\frac{d_{\rm gy}Z^{\alpha}}{dt} \;\equiv\; J^{\alpha\beta}\;\pd{H_{\rm gy}}{Z^{\beta}} \;=\; \{ Z^{\alpha},\; H_{\rm gy}\}_{\rm gc} \;=\;
\frac{1}{B_{\|}^{*}}\,\pd{}{Z^{\beta}}\left( B_{\|}^{*}\,J^{\alpha\beta}\frac{}{} H_{\rm gy} \right),
\label{eq:gy_Ham_eq}
\end{equation}
where we have used the Liouville property of the guiding-center Poisson matrix $J^{\alpha\beta}$ and the guiding-center Jacobian $B_{\|}^{*}$, so that
$\partial_{\alpha}(B_{\|}^{*}\,d_{\rm gy}Z^{\alpha}/dt) \equiv 0$. In addition, the gyrocenter noncanonical Hamilton equations for $({\bf X}, p_{\|})$ are
\begin{eqnarray}
\frac{d_{\rm gy}{\bf X}}{dt} & = & \frac{p_{\|}}{m}\;\frac{{\bf B}^{*}}{B_{\|}^{*}} \;+\; \frac{c\bhat}{eB_{\|}^{*}}\btimes\nabla H_{\rm gy} \;=\;
\frac{p_{\|}}{m}\;\bhat \;+\; \frac{c\bhat}{eB_{\|}^{*}}\btimes\left( \nabla H_{\rm gy} \;+\; \frac{p_{\|}^{2}}{m}\;\bhat\bdot\nabla\bhat \right), 
\label{eq:X_dot} \\
\frac{d_{\rm gy}p_{\|}}{dt} & = & -\;\frac{{\bf B}^{*}}{B_{\|}^{*}}\bdot\nabla H_{\rm gy} \;=\; -\;\bhat\bdot\nabla H_{\rm gy} \;+\; p_{\|}\;\left(\bhat\bdot\nabla\bhat\right)\bdot\frac{d_{\rm gy}{\bf X}}{dt}. 
\label{eq:p_dot}
\end{eqnarray}
The gyrocenter velocity \eqref{eq:X_dot} contains the unperturbed guiding-center velocity $p_{\|}\,({\bf B}^{*}/B_{\|}^{*}) + (c\bhat/eB_{\|}^{*})\btimes \mu\,\nabla B$ as well as the nonlinear perturbed gyrocenter $E\times B$ velocity $(c\bhat/B_{\|}^{*})\btimes\nabla\Phi_{\rm gy}$. The gyrocenter parallel force \eqref{eq:p_dot}, on the other hand, contains the unperturbed guiding-center parallel force $-\,\mu\,({\bf B}^{*}/B_{\|}^{*})\bdot\nabla 
B$ as well as the nonlinear perturbed gyrocenter parallel force $-\,e\,({\bf B}^{*}/B_{\|}^{*})\bdot\nabla\Phi_{\rm gy}$.

Lastly, the gyrocenter magnetic moment $\mu$ is an invariant of the gyrocenter Hamiltonian dynamics,
\begin{equation}
\frac{d_{\rm gy}\mu}{dt} \;=\; -\;\frac{\Omega}{B}\;\pd{H_{\rm gy}}{\zeta} \;\equiv\; 0, 
\label{eq:mu_dot}
\end{equation}
since the gyrocenter Hamiltonian \eqref{eq:H_gy} is independent of the gyrocenter gyroangle, while the gyrocenter gyrofrequency 
\begin{equation}
\frac{d_{\rm gy}\zeta}{dt} \;=\; \frac{\Omega}{B}\;\pd{H_{\rm gy}}{\mu} \;=\; \Omega \;+\; \frac{e\Omega}{B}\;\pd{\Phi_{\rm gy}}{\mu}
\label{eq:zeta_dot}
\end{equation}
contains corrections to the unperturbed gyrofrequency $\Omega$ that are due to the fluctuating electric field.

\subsubsection{Gyrocenter canonical Hamilton equations}

In Sec.~\ref{sec:gyro_mom}, we will need to express the gyrocenter Hamilton equations in canonical form which, with the help of the canonical gyrocenter momentum
\begin{equation}
{\bf p}_{\rm gy} \;=\; \frac{e}{c}\,{\bf A} \;+\; p_{\|}\,\bhat \;\equiv\; \frac{e}{c}\;{\bf A}^{*},
\label{eq:pgy_can}
\end{equation}
satisfies the canonical Hamilton equation
\begin{equation}
\frac{d_{\rm gy}{\bf p}_{\rm gy}}{dt} \;=\; -\;\pd{H_{\rm cgy}}{{\bf X}_{\rm c}},
\label{eq:p_can_Ham}
\end{equation}
where the spatial gradient $\partial/\partial{\bf X}_{\rm c}$ is evaluated at constant ${\bf p}_{\rm gy}$ and $\mu$, with $H_{\rm cgy}({\bf X}_{\rm c},
\mu,{\bf p}_{\rm gy}; t) \equiv H_{\rm gy}({\bf X},p_{\|},\mu; t)$ and ${\bf X}_{\rm c} \equiv {\bf X}$. Equation \eqref{eq:p_can_Ham} is obtained as an Euler-Lagrange equation from the canonical one-form $\Gamma_{\rm cgy} \equiv {\bf p}_{\rm gy}\bdot\exd{\bf X}_{\rm c} + J\,\exd\zeta - H_{\rm cgy}\,\exd 
t$, where the gyroaction coordinate $J \equiv (mc/e)\,\mu$ is canonically conjugate to the gyroangle $\zeta$. 

A direct application of the noncanonical Hamilton equations \eqref{eq:X_dot}-\eqref{eq:p_dot} on the canonical gyrocenter momentum \eqref{eq:pgy_can} yields the Hamiltonian identity
\begin{eqnarray}
\frac{d_{\rm gy}{\bf p}_{\rm gy}}{dt} & = & \frac{e}{c}\;\frac{d_{\rm gy}{\bf X}}{dt}\bdot\nabla{\bf A}^{*} \;+\; \frac{d_{\rm gy}p_{\|}}{dt}\;\bhat 
\;=\; \frac{e}{c} \left( {\bf B}^{*}\btimes\frac{d_{\rm gy}{\bf X}}{dt} \;+\; \nabla{\bf A}^{*}\bdot\frac{d_{\rm gy}{\bf X}}{dt} \right) \;-\;
\left( \frac{{\bf B}^{*}}{B_{\|}^{*}}\bdot\nabla H_{\rm gy}\right)\;\bhat \nonumber \\
 & = & -\;\nabla H_{\rm gy} \;+\; \frac{e}{c}\; \nabla{\bf A}^{*}\bdot\frac{d_{\rm gy}{\bf X}}{dt} \;\equiv\; -\;\pd{H_{\rm cgy}}{{\bf X}_{\rm c}},
\label{eq:pgy_noncan_Ham}
\end{eqnarray}
which relates the two different spatial gradients used in the Sec.~\ref{sec:gyro_mom} [see Eq.~\eqref{eq:Noether_varphi_id}]. We henceforth use the notation $(e/c)\,{\bf A}^{*}$ defined in Eq.~\eqref{eq:pgy_can} whenever we use ${\bf p}_{\rm gy}$ at constant $p_{\|}$ (e.g., $d{\bf p}_{\rm gy} \equiv (e/c)\,d{\bf X}\bdot\nabla{\bf A}^{*} + dp_{\|}\,\bhat$).

In Sec.~\ref{sec:apply}, we will also need to write Eq.~\eqref{eq:p_can_Ham} in axisymmetric tokamak geometry, where the background magnetic field 
\begin{equation}
{\bf B} \;\equiv\; \nabla\varphi\btimes\nabla\psi \;+\; q(\psi)\;\nabla\psi\btimes\nabla\vartheta \;\equiv\; \nabla\btimes{\bf A}, 
\label{eq:B_tok}
\end{equation}
is represented in terms of the standard magnetic coordinates $(\psi,\vartheta,\varphi)$. Here, the {\it safety} factor $q(\psi) \equiv {\bf B}\bdot\nabla\varphi/({\bf B}\bdot\nabla\vartheta)$ is a function of the poloidal flux $\psi$ (note that ${\bf B}\bdot\nabla\psi \equiv 0$), and the vector potential associated with Eq.~\eqref{eq:B_tok} is ${\bf A} \equiv -\,\psi\,\nabla\varphi + A_{\vartheta}(\psi)\;\nabla\vartheta$. From Eq.~\eqref{eq:pgy_can}, the toroidal canonical gyrocenter momentum is therefore defined as
\begin{equation}
p_{{\rm gy}\varphi} \;\equiv\; \pd{\bf X}{\varphi}\bdot{\bf p}_{\rm gy} \;=\; -\,\frac{e}{c}\;\psi \;+\; p_{\|}\;b_{\varphi},
\label{eq:pgy_phi}
\end{equation}
where $b_{\varphi} \equiv \bhat\bdot\partial{\bf X}/\partial\varphi \simeq {\cal R} \equiv |\partial{\bf X}/\partial\varphi|$ is approximately given by the major radius ${\cal R}$ of the axisymmetric tokamak plasma. The toroidal canonical gyrocenter momentum satisfies the canonical Hamilton equation
\begin{equation}
\frac{d_{\rm gy}p_{{\rm gy}\varphi}}{dt} \;=\; \frac{d_{\rm gy}p_{\|}}{dt}\;b_{\varphi} \;-\; \left( \frac{e}{c}\;\frac{d_{\rm gy}\psi}{dt} \;-\; 
p_{\|}\;\frac{d_{\rm gy}b_{\varphi}}{dt} \right) \;=\; -\;\pd{H_{\rm gy}}{\varphi} \;\equiv\; -\,e\;\pd{\Phi_{\rm gy}}{\varphi},
\label{eq:pgy_phi_dot}
\end{equation}
where the last equality follows from the condition of axisymmetry on the background magnetic field. Equation \eqref{eq:pgy_phi_dot} can also be obtained from the noncanonical Euler-Lagrange equation \eqref{eq:EL_X} after using the identity
\begin{eqnarray}
\frac{e}{c}\;\pd{\bf X}{\varphi}\bdot\frac{d_{\rm gy}{\bf X}}{dt}\btimes{\bf B}^{*} & = & \frac{e}{c}{\cal J} \left( \frac{d_{\rm gy}\psi}{dt}\;\nabla\vartheta \;-\; \frac{d_{\rm gy}\vartheta}{dt}\;\nabla\psi \right)\bdot{\bf B}^{*} \nonumber \\
 & = & \frac{e}{c}\;\frac{d_{\rm gy}\psi}{dt} \;-\; p_{\|}\;\frac{d_{\rm gy}b_{\varphi}}{dt},
\label{eq:EL_phi}
\end{eqnarray}
where ${\cal J}$ denotes the Jacobian for the transformation ${\bf X} \rightarrow (\psi,\vartheta,\varphi)$, which is defined from the identity 
\begin{equation}
{\cal J}^{-1} \;\equiv\; (\nabla\psi\btimes\nabla\vartheta)\bdot\nabla\varphi \;=\; B^{\vartheta}. 
\label{eq:Jac_tok}
\end{equation}
Equation \eqref{eq:EL_phi} follows from the identities
\[ \left( \pd{\bf X}{\varphi}\btimes\pd{\bf X}{\psi},\; \pd{\bf X}{\varphi}\btimes\pd{\bf X}{\vartheta}\right) \;\equiv\; \left( {\cal J}\,\nabla\vartheta,
\frac{}{} -\,{\cal J}\,\nabla\psi \right), \] 
with the magnetic-curvature terms
\begin{equation}
\left({\cal J}\,\nabla\vartheta\bdot{\bf B}^{*},\frac{}{} {\cal J}\,\nabla\psi\bdot{\bf B}^{*} \right) \;\equiv\; \left( 1 - \frac{cp_{\|}}{e}\,
\pd{b_{\varphi}}{\psi},\; \frac{cp_{\|}}{e}\,\pd{b_{\varphi}}{\vartheta} \right), 
\label{eq:mag_curv}
\end{equation}
which hold for the axisymmetric tokamak magnetic field \eqref{eq:B_tok}. These identities are then used to construct $d_{\rm gy}b_{\varphi}/dt = 
(\partial b_{\varphi}/\partial\psi)\,d_{\rm gy}\psi/dt + (\partial b_{\varphi}/\partial\vartheta)\,d_{\rm gy}\vartheta/dt$ in Eq.~\eqref{eq:EL_phi}. 

Before moving on to the variational formulation of the gyrokinetic Vlasov-Poisson equations, we introduce the following gyrocenter equations, obtained from the noncanonical gyrocenter Hamilton equations \eqref{eq:X_dot}-\eqref{eq:p_dot}, that will appear in gyrokinetic Vlasov-moment equations presented in Secs.~\ref{subsec:gyro_par} and \ref{subsec:par_tor}. First, the gyrocenter equation for the parallel-momentum vector $p_{\|}\,\bhat$ is
\begin{eqnarray} 
\frac{d_{\rm gy}(p_{\|}\bhat)}{dt} & = & -\;\left(\frac{{\bf B}^{*}}{B_{\|}^{*}}\bdot\nabla H_{\rm gy} \right)\;\bhat \;+\; p_{\|}\;
\frac{d_{\rm gy}\bhat}{dt} \;=\; -\;\nabla H_{\rm gy} \;+\; \frac{e}{c}\;\frac{d_{\rm gy}{\bf X}}{dt}\btimes{\bf B}^{*} \;+\; p_{\|}\;
\frac{d_{\rm gy}\bhat}{dt} \nonumber \\
 & = & -\;\nabla H_{\rm gy} \;+\; \frac{e}{c}\;\frac{d_{\rm gy}{\bf X}}{dt}\btimes{\bf B} \;+\; p_{\|}\;\nabla\bhat\bdot 
\frac{d_{\rm gy}{\bf X}}{dt},
\label{eq:pbhat_dot}
\end{eqnarray}
and, second, the gyrocenter equation for the toroidal component of the parallel-momentum vector $p_{\|}\,b_{\varphi}$ is
\begin{equation}
\frac{d_{\rm gy}}{dt}(p_{\|}\,b_{\varphi}) \;=\; e\,\left( \frac{1}{c}\,\frac{d_{\rm gy}\psi}{dt} \;-\; \pd{\Phi_{\rm gy}}{\varphi} \right) \;\equiv\;
e\,\left( {\bf E}_{\rm gy} \;+\; \frac{1}{c}\,\frac{d_{\rm gy}{\bf X}}{dt}\btimes{\bf B} \right)\bdot\pd{\bf X}{\varphi},
\label{eq:pb_phi_dot}
\end{equation}
where ${\bf E}_{\rm gy} \equiv -\,\nabla\Phi_{\rm gy}$ denotes the generalized gyrocenter electric field and we used the magnetic-coordinate identity
\begin{equation}
\nabla\psi \;\equiv\; {\bf B}\btimes\pd{\bf X}{\varphi}, 
\label{eq:tok_id}
\end{equation}
satisfied by the axisymmetric tokamak magnetic field \eqref{eq:B_tok}. 

Lastly, we introduce an important identity associated with the connection between the canonical gyrocenter momentum Hamilton equation 
\eqref{eq:p_can_Ham} and the canonical gyrocenter toroidal angular-momentum equation \eqref{eq:pgy_phi_dot}. First, we take the dot product of 
Eq.~\eqref{eq:pgy_noncan_Ham} with $\partial{\bf X}/\partial\varphi$ and obtain
\begin{eqnarray}
\frac{d_{\rm gy}{\bf p}_{\rm gy}}{dt}\bdot\pd{\bf X}{\varphi} & = & \frac{d_{\rm gy}p_{{\rm gy}\varphi}}{dt} \;-\; \frac{e}{c}\;
\frac{d_{\rm gy}{\bf X}}{dt}\bdot\nabla\left(\pd{\bf X}{\varphi}\right)\bdot{\bf A}^{*} \nonumber \\
 & = & -\;\pd{H_{\rm gy}}{\varphi} \;+\; \frac{e}{c}\;\pd{{\bf A}^{*}}{\varphi}\bdot\frac{d_{\rm gy}{\bf X}}{dt},
\label{eq:pgy_dot_varphi}
\end{eqnarray}
where the cylindrical dyadic tensor
\begin{equation}
\nabla\left(\pd{\bf X}{\varphi}\right) \;\equiv\; \wh{\cal R}\,\wh{\varphi} \;-\; \wh{\varphi}\,\wh{\cal R}
\label{eq:dyadic}
\end{equation}
is defined in terms of the unit vectors $\wh{\cal R} \equiv \nabla{\cal R}$ and $\wh{\varphi} \equiv {\cal R}^{-1}\,\partial{\bf X}/\partial\varphi$.
Next, we insert Eq.~\eqref{eq:pgy_phi_dot} in Eq.~\eqref{eq:pgy_dot_varphi} and obtain the gyrocenter canonical identity
\begin{equation}
0 \;=\; \frac{d_{\rm gy}p_{{\rm gy}\varphi}}{dt} \;+\; \pd{H_{\rm gy}}{\varphi} \;\equiv\; \frac{e}{c}\,\frac{d_{\rm gy}{\bf X}}{dt}\bdot \left[ 
\pd{{\bf A}^{*}}{\varphi} \;+\; \nabla\left(\pd{\bf X}{\varphi}\right)\bdot{\bf A}^{*} \right].
\label{eq:varphi_id}
\end{equation}
By inserting Eq.~\eqref{eq:dyadic} into Eq.~\eqref{eq:varphi_id}, with $\wh{\sf z} \equiv \wh{\cal R}\btimes\wh{\varphi}$, we obtain the identity
\begin{equation}
\pd{{\bf A}^{*}}{\varphi} \;+\; \nabla\left(\pd{\bf X}{\varphi}\right)\bdot{\bf A}^{*} \;=\; \pd{{\bf A}^{*}}{\varphi} \;+\; {\bf A}^{*}\btimes\wh{\sf z} \;\equiv\; 0,
\label{eq:Astar_varphi}
\end{equation}
which is trivially satisfied in the cylindrical representation ${\bf A}^{*} = A_{\cal R}^{*}\,\wh{\cal R} + A_{\varphi}^{*}\,\wh{\varphi} + A_{\sf z}^{*}\,\wh{\sf z}$, where the components are independent of the toroidal angle $\varphi$ while the unit vectors $\wh{\sf e}^{i} = (\wh{\cal R}, \wh{\varphi},\wh{\sf z})$ satisfy the identities $\partial\wh{\sf e}^{i}/\partial\varphi + \wh{\sf e}^{i}\btimes\wh{\sf z} \equiv 0$.

\subsection{Gyrokinetic Vlasov-Maxwell Variational Principle}

The derivation of exact conservation laws for the nonlinear gyrokinetic equations relies on the existence of variational principles from which they can be derived \cite{Sugama,Brizard_VP,Brizard_gyVP,Brizard_2010}. We now briefly review the variational derivation of the gyrokinetic Vlasov-Poisson equations based on the principle of least action $\delta{\cal A}_{\rm gy} \equiv 0$, where the action functional is \cite{Brizard_gyVP}
\begin{eqnarray}
{\cal A}_{\rm gy} \;=\; \int\;\frac{d^{4}x}{8\pi}\; \left( \epsilon^{2}\,|{\bf E}_{1}|^{2} \;-\frac{}{} |{\bf B}|^{2} \right) \;-\;
\sum\;\int\;d^{8}{\cal Z}\;{\cal F}_{\rm gy}({\cal Z})\;{\cal H}_{\rm gy}({\cal Z};\, \phi_{1}) \;\equiv\; \int\;{\cal L}_{\rm gy}\;d^{4}x,
\label{eq:Agyro_def}
\end{eqnarray}
where the gyrokinetic Lagrangian density ${\cal L}_{\rm gy}$ is the sum of the Maxwell Lagrangian density
\begin{equation}
{\cal L}_{\rm M} \;\equiv\; \frac{1}{8\pi}\;\left(\epsilon^{2}\,|{\bf E}_{1}|^{2} \;-\frac{}{} |{\bf B}|^{2}\right),
\label{eq:L_M_def}
\end{equation}
where the perturbed electric field is ${\bf E}_{1} \equiv -\,\nabla\phi_{1}$, and the gyrocenter Vlasov Lagrangian density
\begin{equation}
{\cal L}_{\rm gyV} \;\equiv\; -\;\sum\;\int\;{\cal F}_{\rm gy}({\cal Z})\;{\cal H}_{\rm gy}({\cal Z};\, \phi_{1})\;d^{4}p,
\label{eq:L_V_def}
\end{equation}
where the extended gyrocenter Hamiltonian
\begin{equation}
{\cal H}_{\rm gy}({\cal Z};\, \phi_{1}) \;\equiv\; H_{\rm gy}({\bf X}, p_{\|}, \mu, t;\; \phi_{1}) \;-\; W
\label{eq:Hext_def}
\end{equation}
is expressed in terms of the (regular) time-dependent gyrocenter Hamiltonian \eqref{eq:H_gy} and the gyrocenter energy coordinate $W$. The extended phase-space integration in Eq.~\eqref{eq:Agyro_def} is defined with $d^{8}{\cal Z} \equiv dt\,d^{3}X\; d^{4}p$, where $d^{4}p \equiv 
c^{-1}dW\;d^{3}p$ and $d^{3}p = mB_{\|}^{*}\;dp_{\|}\,d\mu\,d\zeta$ (here, $m\,B_{\|}^{*}$ denotes the Jacobian). Furthermore, the extended gyrocenter Vlasov distribution
\begin{equation}
{\cal F}_{\rm gy}({\cal Z}) \;\equiv\; c\,\delta(W - H_{\rm gy})\; F({\bf X}, p_{\|}, \mu, t)
\label{eq:Fext_def}
\end{equation}
ensures that the gyrocenter Hamiltonian dynamics satisfies the physical constraint ${\cal H}_{\rm gy} \equiv 0$. 

\subsubsection{Eulerian variations}

The gyrokinetic Vlasov-Poisson equations describing the self-consistent evolution of the gyrocenter Vlasov distribution $F$ in the presence of electrostatic fluctuations $\phi_{1}$ are obtained from the Eulerian variational principle
\begin{equation}
\int\;\delta{\cal L}_{\rm gy}\;d^{4}x \;=\; 0,
\label{eq:gyro_vp}
\end{equation}
where the Eulerian variation of the gyrokinetic Lagrangian density is expressed as
\begin{eqnarray}
\delta{\cal L}_{\rm gy} & = & \frac{\epsilon^{2}}{4\pi}\;\left( \delta{\bf E}_{1}\bdot{\bf E}_{1} \right) \;-\; \sum\;\int \left[\; 
\delta{\cal F}_{\rm gy}\,{\cal H}_{\rm gy} + \epsilon\,\delta\phi_{1} \left( \epsilon^{-1}\fd{H_{\rm gy}}{\phi_{1}} \right)\;{\cal F}_{\rm gy} \;\right] 
d^{4}p,
\label{eq:delta_L_def} 
\end{eqnarray}
with $d^{3}p = 2\pi\,mB_{\|}^{*}\;dp_{\|}\,d\mu$ (which henceforth assumes gyroangle-averaging). In addition, ${\bf B}$ is not a variational field (since it is time independent) while the Eulerian variations $\delta{\bf E}_{1} \equiv -\,\nabla\delta\phi_{1}$ preserves the electrostatic constraint $\nabla\btimes\delta{\bf E}_{1} = 0$. The Eulerian variation for the extended gyrocenter distribution \eqref{eq:Fext_def} is $\delta{\cal F}_{\rm gy} \equiv \{ {\cal S}_{\rm gy},\; {\cal F}_{\rm gy}\}_{\rm gc}$, which preserves the Vlasov constraint $\int\delta{\cal F}_{\rm gy}\,d^{8}{\cal Z} \equiv 0$ under a virtual canonical transformation ${\cal Z}^{a} \rightarrow {\cal Z}^{a} + \delta{\cal Z}^{a}$ in extended phase space, where $\delta{\cal Z}^{a} \equiv \{ {\cal Z}^{a},\; {\cal S}_{\rm gy}\}_{\rm gc}$ is generated by the extended scalar field 
${\cal S}_{\rm gy}$.

Equations \eqref{eq:H_gy} and \eqref{eq:S1_def} yield the functional derivative
\begin{equation}
\epsilon^{-1}\;\fd{H_{\rm gy}}{\phi_{1}({\bf x})} \;\equiv\; e\; \left\langle {\sf T}_{\rm gy}^{-1}\,\delta_{\rm gc}^{3}\right\rangle \;=\; 
e\;\left\langle \delta_{\rm gc}^{3} \;-\frac{}{} \epsilon\,\{ S_{1},\; \delta_{\rm gc}^{3}\}_{\rm gc} \;+\; \cdots \right\rangle,
\label{eq:fdH_Phi}
\end{equation}
where the guiding-center delta function $\delta^{3}_{\rm gc} \equiv \delta^{3}({\bf X} + \vb{\rho}_{\rm gc} - {\bf x})$ indicates that the gyrocenter contribution at a fixed field point ${\bf x}$ only comes from gyrocenters located on the ring ${\bf X} = {\bf x} - \vb{\rho}_{\rm gc}$. It is convenient to formally express the combined guiding-center and gyrocenter push-forward operations
\begin{equation}
{\sf T}_{\rm gy}^{-1}\delta_{\rm gc}^{3} \;\equiv\; {\sf T}_{\epsilon}^{-1}\delta^{3}({\bf X} - {\bf x}) \;\equiv\; \delta^{3}({\bf X} + 
\vb{\rho}_{\epsilon} - {\bf x})
\label{eq:delta_epsilon}
\end{equation}
in terms of the reduced displacement
\begin{equation}
\vb{\rho}_{\epsilon} \;\equiv\; \vb{\rho}_{0} \;+\; \vb{\rho}_{1} + \cdots,
\label{eq:rho_epsilon}
\end{equation}
where the unperturbed gyrocenter displacement $\vb{\rho}_{0} \equiv \vb{\rho}_{\rm gc0}$ denotes the (lowest-order) guiding-center gyroradius (which is
gyroangle-dependent) and the first-order reduced displacement 
\begin{equation}
\vb{\rho}_{1} \;\equiv\; \epsilon_{B}\,\vb{\rho}_{\rm gc 1} \;+\; \epsilon\,\vb{\rho}_{\rm gy 1} \;=\; \langle\vb{\rho}_{1}\rangle \;+\; 
\wt{\vb{\rho}}_{1}
\label{eq:rho_1_total}
\end{equation}
includes the first-order guiding-center displacement $\vb{\rho}_{\rm gc1}$ \cite{RGL,Cary_Briz} and the first-order gyrocenter displacement
\begin{equation}
\vb{\rho}_{\rm gy 1} \;\equiv\; \left\{ {\bf X} \;+\; \vb{\rho}_{\rm gc},\frac{}{} S_{1} \right\}_{\rm gc},
\label{eq:rhogy_def}
\end{equation}
which is expressed exclusively in terms of $S_{1}$ (and therefore $\wt{\phi}_{1{\rm gc}}$) according to Eq.~\eqref{eq:S1_def}. Each first-order contribution in Eq.~\eqref{eq:rho_1_total} has a gyroangle-independent part, which contributes an electric-dipole term in the gyrokinetic polarization [see Eqs.~\eqref{eq:pol_epsilon}-\eqref{eq:rho1_gy}], and a gyroangle-dependent part, which contributes an electric-quadrupole term in the gyrokinetic polarization [see Eq.~\eqref{eq:quad_def}]. (We note that effects due to the gyrokinetic magnetization \cite{Brizard_2010} do not enter into our present discussion but will be considered in future work.)

By rearranging terms in the Eulerian variation \eqref{eq:delta_L_def} so that the variation generators $({\cal S}_{\rm gy}, \delta\phi_{1})$ appear isolated, we obtain
\begin{eqnarray}
\delta{\cal L}_{\rm gy} & = & \epsilon\,\delta\phi_{1} \left[ \frac{\epsilon}{4\pi}\;\nabla\bdot{\bf E}_{1} \;-\; \sum\;e\; \int\;{\cal F}_{\rm gy}\; \left\langle {\sf T}_{\rm gy}^{-1}\,\delta_{\rm gc}^{3}\right\rangle\;d^{4}p\;d^{3}X \;\right] \nonumber \\
 &  &-\; \sum\;\int\; {\cal S}_{\rm gy}\; \left\{ {\cal F}_{\rm gy},\frac{}{} {\cal H}_{\rm gy} \right\}_{\rm gc}\;d^{4}p \;+\; \left( \pd{\Lambda}{t} 
\;+\; \nabla\bdot\vb{\Gamma}\right),
\label{eq:delta_L}
\end{eqnarray}
where the Noether fields $\Lambda$ and $\vb{\Gamma}$ are
\begin{eqnarray}
\Lambda & \equiv & \sum\;\int\; {\cal S}_{\rm gy}\;{\cal F}_{\rm gy}\;d^{4}p, \label{eq:Lambda_def} \\
\vb{\Gamma} & \equiv & -\; \frac{\epsilon^{2}}{4\pi} \;\delta\phi_{1}\;{\bf E}_{1} \;+\; \sum\;\int\; \left({\cal S}_{\rm gy}\frac{}{}{\cal F}_{\rm gy}\right)\;\frac{d_{\rm gy}{\bf X}}{dt}\;d^{4}p. 
\label{eq:Gamma_def}
\end{eqnarray}
We note that the Noether space-time divergence terms $\partial\Lambda/\partial t + \nabla\bdot\vb{\Gamma}$ do not contribute to the Eulerian variational principle \eqref{eq:gyro_vp} but instead play a crucial role in the derivation of exact conservation laws by the Noether method (see Sec.~\ref{sec:gyro_mom}).

\subsubsection{Gyrokinetic Vlasov-Poisson equations}

When the variation \eqref{eq:delta_L} is inserted into the Eulerian variational principle \eqref{eq:gyro_vp} for arbitrary variation generators 
$({\cal S}_{\rm gy}, \delta\phi_{1})$, we obtain the gyrokinetic Vlasov equation in extended phase space
\begin{equation}
\{ {\cal F}_{\rm gy},\; {\cal H}_{\rm gy} \}_{\rm gc} \;=\; 0,
\label{eq:ext_Veq}
\end{equation}
and the gyrokinetic Poisson equation
\begin{equation}
\epsilon\;\nabla\bdot{\bf E}_{1} \;\equiv\; 4\pi\,\sum\;e\; \int\;{\cal F}_{\rm gy}\; \left\langle {\sf T}_{\rm gy}^{-1}\,\delta_{\rm gc}^{3}\right\rangle\;d^{4}p\;d^{3}X. 
\label{eq:gyro_E} 
\end{equation}
Furthermore, when the energy integration $(\int dW)$ is performed on the extended gyrokinetic Vlasov equation \eqref{eq:ext_Veq}, we obtain the gyrokinetic Vlasov equation
\begin{eqnarray}
0 & = & \pd{F}{t} \;+\; \left\{ F,\frac{}{} H_{\rm gy}\right\}_{\rm gc} \;\equiv\; \pd{F}{t} \;+\; \frac{d_{\rm gy}{\bf X}}{dt}\bdot\nabla F \;+\; 
\frac{d_{\rm gy}p_{\|}}{dt}\;
\pd{F}{p_{\|}},
\label{eq:gyro_V}
\end{eqnarray}
where $\partial F/\partial\mu$ is absent because of Eq.~\eqref{eq:mu_dot} and the gyrocenter gyrofrequency \eqref{eq:zeta_dot} does not appear in 
Eq.~\eqref{eq:gyro_V} since $\partial F/\partial\zeta \equiv 0$ (by construction). Hence, the guiding-center and gyrocenter transformations yield a reduced dynamical Hamiltonian description of gyrocenter motion in terms of four-dimensional coordinates ${\bf X}$ and $p_{\|}$, with $\mu$ appearing as a parameter.

The gyrokinetic Poisson equation \eqref{eq:gyro_E}, using ${\bf E}_{1} \equiv -\,\nabla\phi_{1}$, becomes
\begin{eqnarray}
-\;\epsilon\;\nabla^{2}\phi_{1} & = & 4\pi\,\sum\;e\; \int\;F\; \left\langle {\sf T}_{\rm gy}^{-1}\,\delta_{\rm gc}^{3}\right\rangle\;d^{6}z 
\nonumber \\
 & \equiv & 4\pi\,\left( \varrho_{\rm gy} \;-\frac{}{} \nabla\bdot\vb{\cal P}\right),
\label{eq:gyroMax_E}
\end{eqnarray}
where the gyrocenter charge density is defined as
\begin{equation}
\varrho_{\rm gy} \;\equiv\; \sum\;e\;\int\;F\;d^{3}p,
\label{eq:rho_gy}
\end{equation}
and the gyrokinetic polarization
\begin{equation}
\vb{\cal P} \;\equiv\; \sum\;e\;\int\;F\;\langle\vb{\rho}_{\epsilon}\rangle\;d^{3}p \;-\; \nabla\bdot{\sf Q}
\label{eq:pol_epsilon}
\end{equation}
includes the electric-dipole contribution associated with the first-order reduced displacement \eqref{eq:rho_epsilon}, where $\langle
\vb{\rho}_{\epsilon}\rangle \equiv \langle\vb{\rho}_{1}\rangle + \cdots$ (since $\langle\vb{\rho}_{0}\rangle \equiv 0$), as well as a quadrupole contribution represented by the second-rank tensor
\begin{equation}
{\sf Q} \;\equiv\; \sum\;\frac{e}{2}\;\int\; F \;\left\langle\vb{\rho}_{\epsilon}\frac{}{}\vb{\rho}_{\epsilon}\right\rangle d^{3}p \;+\; \cdots,
\label{eq:quad_def}
\end{equation}
which includes standard finite-Larmor-radius corrections due to $\langle\vb{\rho}_{0}\vb{\rho}_{0}\rangle$ as well as the first-order corrections
$\langle\vb{\rho}_{0}\,\wt{\vb{\rho}}_{1}\rangle$ and $\langle\wt{\vb{\rho}}_{1}\,\vb{\rho}_{0}\rangle$ associated with the gyroangle-dependent part of the first-order displacement \eqref{eq:rho_1_total}. The first term in the gyrokinetic polarization \eqref{eq:pol_epsilon} includes the first-order guiding-center contribution \cite{Kaufman_86}
\begin{equation}
\langle\vb{\rho}_{\rm gc 1}\rangle \;\equiv\; \frac{\bhat}{\Omega}\btimes\frac{d_{\rm gc}{\bf X}}{dt} \;=\; -\;\frac{1}{m\Omega^{2}} \left( \mu\,\nabla_{\bot}B \;+\; \frac{p_{\|}^{2}}{m}\;\bhat\bdot\nabla\bhat \right),
\label{eq:rho1_gc}
\end{equation}
and the first-order gyrocenter contribution
\begin{equation}
\langle\vb{\rho}_{\rm gy 1}\rangle \;\equiv\; -\;\frac{e}{B}\;\pd{}{\mu}\left\langle \vb{\rho}_{0}\frac{}{}\wt{\phi}_{1{\rm gc}}\right\rangle \;=\; -\;\frac{c}{B\Omega}\;\nabla_{\bot}\langle\phi_{1{\rm gc}}\rangle.
\label{eq:rho1_gy}
\end{equation}
By using Eqs.~\eqref{eq:rho1_gc}-\eqref{eq:rho1_gy}, the gyrokinetic polarization \eqref{eq:pol_epsilon} can thus be expressed in the form
\begin{equation}
\vb{\cal P} \;=\; \sum\;m\,\frac{c\bhat}{B}\btimes\left[ \int\; F\;\left(\frac{d_{\rm gy}^{(1)}{\bf X}}{dt}\right)_{\bot}\; d^{3}p \right] \;-\; 
\nabla\bdot{\sf Q},
\label{eq:pol_dipole}
\end{equation}
where the truncated gyrocenter velocity
\begin{equation}
\left(\frac{d_{\rm gy}^{(1)}{\bf X}}{dt}\right)_{\bot} \;\equiv\; \frac{c\bhat}{e\,B}\btimes\left[\; \epsilon_{B}\;
\left( \mu\;\nabla B \;+\; \frac{p_{\|}^{2}}{m}\;\bhat\bdot\nabla\bhat\right) \;+\; \epsilon\,e\;\nabla\langle\phi_{1{\rm gc}}\rangle \;\right]
\label{eq:dgyX_dt_1}
\end{equation}
denotes the perpendicular components of the gyrocenter velocity \eqref{eq:X_dot} with the effective gyrocenter
potential \eqref{eq:Psi_gy_def} replaced with its first-order contribution $\epsilon\,\langle\phi_{1{\rm gc}}\rangle$.

\section{\label{sec:gyro_mom}Gyrokinetic Momentum Conservation Laws}

In this Section, we use the Noether method \cite{Sugama,Brizard_VP,Brizard_gyVP,Brizard_Hahm,Brizard_2010} to derive an exact momentum conservation law for the gyrokinetic Vlasov-Poisson equations \eqref{eq:gyro_V}-\eqref{eq:gyroMax_E}. After substituting the gyrokinetic Vlasov-Poisson equations \eqref{eq:ext_Veq}-\eqref{eq:gyro_E} into the Eulerian variational equation \eqref{eq:delta_L}, we obtain the gyrokinetic Noether equation
\begin{equation}
\delta{\cal L}_{\rm gy} \;=\; \pd{\Lambda}{t} \;+\; \nabla\bdot\vb{\Gamma},
\label{eq:gyro_Noether}
\end{equation}
where the variations $({\cal S}_{\rm gy}, \delta\phi_{1})$ in Eqs.~\eqref{eq:Lambda_def}-\eqref{eq:Gamma_def} are no longer considered arbitrary but are instead generated by infinitesimal space-time translations or rotations. The energy conservation law for the gyrokinetic Vlasov-Maxwell equations was obtained from the gyrokinetic Noether equation \eqref{eq:gyro_Noether} and was discussed in Refs.~\cite{Brizard_1989a,Brizard_1989b,Sugama,Brizard_VP,Brizard_gyVP,Brizard_Hahm,Brizard_2010} by considering infinitesimal time translations $t \rightarrow t + \delta t$.

\subsection{Noether Method}

The momentum conservation law for the gyrokinetic Vlasov-Poisson equations \eqref{eq:gyro_V}-\eqref{eq:gyroMax_E} is obtained from the gyrokinetic Noether equation \eqref{eq:gyro_Noether} by considering infinitesimal space translations ${\bf x} \rightarrow {\bf x} + \delta {\bf x}$ 
\cite{Brizard_VP,Brizard_JPP,Brizard_JPCS}, for which 
\begin{equation}
\left. \begin{array}{rcl}
{\cal S}_{\rm gy} & = & {\bf p}_{\rm gy}\bdot\delta{\bf x} \\
 && \\
\delta\phi_{1} & = & -\;\delta{\bf x}\bdot\nabla\phi_{1} \;\equiv\; \delta{\bf x}\bdot{\bf E}_{1} \\
 && \\
\delta{\cal L}_{\rm gy} & = & -\;\nabla\bdot\left(\delta{\bf x}\frac{}{}{\cal L}_{\rm gy}\right) \;+\; \delta{\bf x}\bdot\nabla^{\prime}{\cal L}_{\rm gy}
\end{array} \right\},
\label{eq:delta_x_var}
\end{equation}
where the gyrocenter generating scalar field ${\cal S}_{\rm gy}$ generates a virtual spatial translation $\delta{\bf x}$ and the gyrocenter canonical momentum ${\bf p}_{\rm gy}$ is defined in Eq.~\eqref{eq:pgy_can}. 

The first term in the expression for $\delta{\cal L}_{\rm gy}$ in Eq.~\eqref{eq:delta_x_var} takes into account the geometric nature of 
${\cal L}_{\rm gy}$ as a space-time density. The notation $\nabla^{\prime}{\cal L}_{\rm gy}$ in the second term, on the other hand, denotes the explicit spatial gradient of the Lagrangian density ${\cal L}_{\rm gy}$ with the gyrokinetic fields $(F, {\bf E}_{1})$ held constant (i.e., $\nabla^{\prime}
\equiv 0$ in a uniform magnetic field). Hence, the explicit gradient $\nabla^{\prime}{\cal L}_{\rm gy}$ becomes
\begin{eqnarray}
\nabla^{\prime}{\cal L}_{\rm gy} & = & -\;\frac{B}{4\pi}\,\nabla B \;-\; \sum\;\int\;F\;\left( \nabla^{\prime}H_{\rm gy} \;-\; 
\frac{e}{c}\,\nabla{\bf A}^{*}\bdot\frac{d_{\rm gy}{\bf X}}{dt} \right)\;d^{3}p,
\label{eq:nabla_prime}
\end{eqnarray}
where the first term represents the contribution from the Maxwell part \eqref{eq:L_M_def} of the Lagrangian density while the gyrocenter Vlasov term involves the explicit canonical gradient expressed in terms of the Hamiltonian identity \eqref{eq:pgy_noncan_Ham}. 

By inserting the variations \eqref{eq:delta_x_var} into the gyrokinetic Noether equation \eqref{eq:gyro_Noether}, we obtain the {\it primitive} gyrokinetic Noether momentum equation
\begin{eqnarray}
\delta{\bf x}\bdot\nabla^{\prime}{\cal L}_{\rm gy} & = & \pd{}{t} \left[\; \sum\;\int\;F\,
\left({\bf p}_{\rm gy}\frac{}{}\bdot\;\delta{\bf x}\right)\;d^{3}p \;\right] 
\nonumber \\
 &  &+\;\nabla\bdot\left[\; \sum\;\int\; F\,\frac{d_{\rm gy}{\bf X}}{dt}\;\left({\bf p}_{\rm gy}\frac{}{}\bdot\;\delta{\bf x}\right)\;d^{3}p \;-\; 
\frac{\epsilon^{2}}{4\pi} \;\left(\delta{\bf x}\bdot\frac{}{} {\bf E}_{1}\right)\;{\bf E}_{1} \;+\; {\cal L}_{\rm M}\;\delta{\bf x} \right],
\label{eq:gyro_mom_prim} 
\end{eqnarray}
where only the Maxwell part \eqref{eq:L_M_def} of the gyrokinetic Lagrangian defined in Eq.~\eqref{eq:Agyro_def} survives the physical constraint 
${\cal H}_{\rm gy} \equiv 0$.  Next, if we replace ${\cal L}_{\rm M}$ and $\nabla^{\prime}{\cal L}_{\rm gy}$ in Eq.~\eqref{eq:gyro_mom_prim} with 
Eqs.~\eqref{eq:L_M_def} and \eqref{eq:nabla_prime}, respectively, we obtain
\begin{eqnarray}
-\;\delta{\bf x}\bdot\left[ \sum\;\int\;F\;\left( \nabla^{\prime}H_{\rm gy} \;-\; 
\frac{e}{c}\,\nabla{\bf A}^{*}\bdot\frac{d_{\rm gy}{\bf X}}{dt} \right)\;d^{3}p \right] & = & \pd{}{t} \left[\; \sum\;\int\;{\cal F}_{\rm gy}\,
\left({\bf p}_{\rm gy}\frac{}{}\bdot\;\delta{\bf x}\right)\;d^{4}p \;\right] \nonumber \\
 &  &+\; \nabla\bdot\left[\; \sum\;\int\; {\cal F}_{\rm gy}\,\frac{d_{\rm gy}{\bf X}}{dt}\;\left({\bf p}_{\rm gy}\frac{}{}\bdot\;\delta{\bf x}\right)\;
d^{4}p \;\right] \nonumber \\
 &  &-\;\nabla\bdot\left[\; \frac{\epsilon^{2}}{4\pi} \;\left(\delta{\bf x}\bdot\frac{}{} {\bf E}_{1}\right)\;{\bf E}_{1} \;+\; 
\left(\frac{\epsilon^{2}}{8\pi}\,|{\bf E}_{1}|^{2}\right)\;\delta{\bf x} \right],
\label{eq:gyro_mom_prime} 
\end{eqnarray}
where the right side involves the gyrocenter canonical momentum ${\bf p}_{\rm gy}$ and the perturbed electric field ${\bf E}_{1}$.

\subsection{\label{subsec:gyro_lin}Gyrokinetic Linear Momentum}

The gyrokinetic Noether (linear) momentum equation is obtained from Eq.~\eqref{eq:gyro_mom_prime} by considering a virtual translation generated by an arbitrary constant displacement $\delta{\bf x}$. The gyrokinetic Noether momentum equation is, therefore, expressed as
\begin{equation}
\pd{{\bf P}}{t} \;+\; \nabla\bdot\vb{\Pi} \;=\; -\sum\int\;F\;\left( \nabla^{\prime}H_{\rm gy} \;-\; \frac{e}{c}\,\nabla{\bf A}^{*}\bdot
\frac{d_{\rm gy}{\bf X}}{dt} \right) \;d^{3}p,
\label{eq:gyro_mom_final}
\end{equation}
where the gyrokinetic momentum density is
\begin{equation}
{\bf P} \;=\; \sum\;\int\; F\,{\bf p}_{\rm gy}\;d^{3}p,
\label{eq:Pi_gyro}
\end{equation}
and the {\it canonical} gyrokinetic momentum-stress tensor is
\begin{equation}
\vb{\Pi} \;=\; \frac{\epsilon^{2}}{4\pi}\;\left( |{\bf E}_{1}|^{2}\;\frac{{\bf I}}{2} \;-\; {\bf E}_{1}\,{\bf E}_{1} \right) \;+\; \sum\;\int \;F\; 
\frac{d_{\rm gy}{\bf X}}{dt}\;{\bf p}_{\rm gy} \;d^{3}p \;\equiv\; \vb{\Pi}_{\rm E} \;+\; \vb{\Pi}_{\rm gy}.
\label{eq:T_gyro}
\end{equation}
Here, the canonical stress tensor contains the symmetric Maxwell stress tensor 
\begin{equation}
\vb{\Pi}_{\rm E} \;\equiv\; \frac{\epsilon^{2}}{4\pi}\;\left( |{\bf E}_{1}|^{2}\;\frac{{\bf I}}{2} \;-\; {\bf E}_{1}\,{\bf E}_{1} \right),
\label{eq:Pi_E_def}
\end{equation}
and the asymmetric gyrocenter canonical momentum-stress tensor
\begin{equation}
\vb{\Pi}_{\rm gy} \;\equiv\; \sum\;\int \;F\; \frac{d_{\rm gy}{\bf X}}{dt}\;{\bf p}_{\rm gy} \;d^{3}p.
\label{eq:Pi_gy_def}
\end{equation}
We note that, while the asymmetry of the gyrocenter canonical momentum-stress tensor \eqref{eq:Pi_gy_def} is not physically relevant here since it appears in the gyrokinetic Vlasov-moment equation \eqref{eq:gyro_moment}, which has no symmetry requirements on $\vb{\Pi}_{\rm gy}$, it plays a crucial role in deriving the exact gyrokinetic toroidal angular-momentum conservation law [see Eqs.~\eqref{eq:angmom_varphi}-\eqref{eq:Noether_varphi_id}].

We now show that Eq.~\eqref{eq:gyro_mom_final} is consistent with Eq.~\eqref{eq:p_can_Ham} by constructing an explicit proof of the gyrokinetic Noether momentum equation \eqref{eq:gyro_mom_final}. We begin with the partial time derivative of the gyrokinetic momentum density \eqref{eq:Pi_gyro}
\begin{equation}
\pd{{\bf P}}{t} \;=\; \sum\;\int \left(\; \pd{F}{t}\;{\bf p}_{\rm gy} \;+\; F \pd{{\bf p}_{\rm gy}}{t}\;\right)\;d^{3}p.
\label{eq:pi_dot}
\end{equation}
Upon substituting the phase-space divergence form of the gyrokinetic Vlasov equation \eqref{eq:gyro_V}, Eq.~\eqref{eq:pi_dot} becomes
\begin{eqnarray}
\pd{{\bf P}}{t} & = & \sum\;\int\;F \;\left(\frac{d_{\rm gy}{\bf p}_{\rm gy}}{dt}\right) \;d^{3}p \;-\; \nabla\bdot\vb{\Pi}_{\rm gy}.
\label{eq:pi_dot_VP}
\end{eqnarray}
Next, we compute the divergence of the canonical momentum-stress tensor \eqref{eq:T_gyro} to obtain 
\begin{eqnarray}
\nabla\bdot\vb{\Pi} & = & -\,\epsilon\;\sum\;e\;\int\; F\;\left\langle{\sf T}_{\rm gy}^{-1}{\bf E}_{1{\rm gc}}\right\rangle \;d^{3}p \;+\; 
\nabla\bdot\vb{\Pi}_{\rm gy},
\label{eq:div_pi_VP}
\end{eqnarray}
where we used the gyrokinetic Poisson equation \eqref{eq:gyro_E} with ${\bf E}_{1} \equiv -\,\nabla\phi_{1}$ and ${\bf E}_{1{\rm gc}} \equiv
-\,\nabla\phi_{1{\rm gc}}$, and we used the Maxwell-tensor divergence
\begin{equation}
\nabla\bdot\vb{\Pi}_{\rm E} \;\equiv\; -\,\epsilon\;\sum\;e\;\int\; F\;\left\langle{\sf T}_{\rm gy}^{-1}{\bf E}_{1{\rm gc}}\right\rangle \;d^{3}p.
\label{eq:div_Pi_E}
\end{equation}
By inserting Eqs.~\eqref{eq:pi_dot_VP}-\eqref{eq:div_pi_VP} into Eq.~\eqref{eq:gyro_mom_final}, we obtain the gyrokinetic momentum equation
\begin{equation}
0 \equiv \sum\int\,F \left[ \frac{d_{\rm gy}{\bf p}_{\rm gy}}{dt} - \epsilon\,e\; \left\langle {\sf T}_{\rm gy}^{-1}
{\bf E}_{1{\rm gc}} \right\rangle \;+\; \nabla^{\prime}H_{\rm gy} \;-\; \frac{e}{c}\,\nabla{\bf A}^{*}\bdot
\frac{d_{\rm gy}{\bf X}}{dt} \right]\;d^{3}p.
\label{eq:linear_mom_test}
\end{equation}
We immediately conclude that the canonical gyrocenter momentum ${\bf p}_{\rm gy}$ must satisfy the gyrocenter equation of motion
\begin{equation}
\frac{d_{\rm gy}{\bf p}_{\rm gy}}{dt} \;\equiv\; \epsilon\,e\; \left\langle {\sf T}_{\rm gy}^{-1}{\bf E}_{1{\rm gc}} \right\rangle 
\;-\; \left( \nabla^{\prime}H_{\rm gy} \;-\; \frac{e}{c}\,\nabla{\bf A}^{*}\bdot\frac{d_{\rm gy}{\bf X}}{dt} \right).
\label{eq:Pi_can_dot}
\end{equation}
We note, however, that the expression for $\langle {\sf T}_{\rm gy}^{-1}{\bf E}_{1{\rm gc}}\rangle$ can be rewritten as
\begin{eqnarray}
\epsilon\;\left\langle {\sf T}_{\rm gy}^{-1}{\bf E}_{1{\rm gc}} \right\rangle & = & -\,\epsilon\,\nabla\langle\phi_{1{\rm gc}}\rangle \;+\; \epsilon^{2}
\left\langle\left\{ S_{1},\frac{}{} \nabla\phi_{1{\rm gc}}\right\} \right\rangle \;+\; \cdots \nonumber \\
 & \equiv & -\,\left( \nabla\Phi_{\rm gy} \;-\frac{}{} \nabla^{\prime}\Phi_{\rm gy}\right),
\label{eq:Tgy_E}
\end{eqnarray}
where the explicit gradient term $\nabla^{\prime}\Phi_{\rm gy}$ takes into account the nonuniformity of the background magnetic field (e.g., which appears in the guiding-center Poisson bracket). Lastly, using the relation $e\,\nabla^{\prime}\Phi_{\rm gy} - \nabla^{\prime}H_{\rm gy} \equiv -\,\nabla 
H_{\rm gc}$, we immediately recover the canonical gyrocenter momentum equation \eqref {eq:p_can_Ham} from Eq.~\eqref{eq:Pi_can_dot}:
\begin{eqnarray}
\frac{d_{\rm gy}{\bf p}_{\rm gy}}{dt} & = & -\,\epsilon\,e\left(\nabla\Phi_{\rm gy} \;-\frac{}{} \nabla^{\prime}\Phi_{\rm gy} \right) \;-\;
\left( \nabla^{\prime}H_{\rm gy} \;-\; \frac{e}{c}\,\nabla{\bf A}^{*}\bdot\frac{d_{\rm gy}{\bf X}}{dt} \right) \nonumber \\
 & = & -\,\epsilon\,e\;\nabla\Phi_{\rm gy} \;-\; \left( \nabla H_{\rm gc} \;-\; \frac{e}{c}\,\nabla{\bf A}^{*}\bdot\frac{d_{\rm gy}{\bf X}}{dt} \right) 
\;\equiv\; -\;\nabla H_{\rm gy} \;+\; \frac{e}{c}\,\nabla{\bf A}^{*}\bdot\frac{d_{\rm gy}{\bf X}}{dt}.
\label{eq:p_can_Ham_Noether}
\end{eqnarray}
We therefore see that the gyrokinetic Noether momentum equation \eqref{eq:gyro_mom_final} is automatically guaranteed by the canonical gyrocenter momentum equation \eqref{eq:p_can_Ham}. Lastly, we note that Eq.~\eqref{eq:linear_mom_test} can be written as
\begin{eqnarray} 
0 & \equiv & \sum\int\,F \left[\frac{d_{\rm gy}{\bf p}_{\rm gy}}{dt} \;+\; \left( \nabla H_{\rm gy} \;-\; \frac{e}{c}\,\nabla{\bf A}^{*}\bdot
\frac{d_{\rm gy}{\bf X}}{dt} \right)\right]\;d^{3}p \nonumber \\
 & = & \pd{{\bf P}}{t} \;+\; \nabla\bdot\vb{\Pi}_{\rm gy} \;+\; \sum\int\,F \;\left( \nabla H_{\rm gy} \;-\; 
\frac{e}{c}\,\nabla{\bf A}^{*}\bdot\frac{d_{\rm gy}{\bf X}}{dt} \right)\;d^{3}p,
\label{eq:gyro_moment}
\end{eqnarray}
which can be recovered as a gyrocenter Vlasov-moment equation 
\begin{equation}
0 \;=\; \pd{}{t}\left( \int\;\chi\frac{}{}F\,d^{3}p\right) \;+\; \nabla\bdot\left(\int\;\frac{d_{\rm gy}{\bf X}}{dt}\;\chi\;F\,d^{3}p\right) \;-\;
\int\;\frac{d_{\rm gy}\chi}{dt}\;F\;d^{3}p
\label{eq:chi_moment}
\end{equation}
obtained from the gyrokinetic Vlasov identity
\begin{equation}
0 \;\equiv\; \pd{}{t}\left(B_{\|}^{*}\,\chi\frac{}{}F\right) \;+\; \pd{}{Z^{\alpha}}\left( B_{\|}^{*}\,\chi\frac{}{}F\;\left\{ Z^{\alpha},
\frac{}{} H_{\rm gy}\right\}_{\rm gc} \right) \;-\; B_{\|}^{*}\,F\;\frac{d_{\rm gy}\chi}{dt},
\label{eq:chi_Vlasov}
\end{equation}
which holds for any function $\chi$ on gyrocenter phase space (e.g., $\chi = {\bf p}_{\rm gy}$). We note that Eq.~\eqref{eq:chi_Vlasov} follows from writing the gyrokinetic Vlasov equation \eqref{eq:gyro_V} in phase-space-divergence form using Eq.~\eqref{eq:gy_Ham_eq}.

\subsection{\label{subsec:gyro_par}Gyrokinetic Parallel Momentum}

As a simple application of the gyrokinetic Noether momentum equation \eqref{eq:gyro_mom_final}, we now derive the gyrokinetic parallel momentum equation by first decomposing the gyrokinetic momentum density \eqref{eq:Pi_gyro} and the gyrocenter momentum-stress tensor \eqref{eq:Pi_gy_def} as
\begin{equation}
\left( {\bf P},\; \vb{\Pi}_{\rm gy} \right) \;=\; \left( {\bf P}_{\|} \;+\; \frac{1}{c}\;\varrho_{\rm gy}\;{\bf A},\; \vb{\Pi}_{{\rm gy}\|} \;+\; 
\frac{1}{c}\;{\bf J}_{\rm gy}\;{\bf A}\right), 
\label{eq:P_rhoJ}
\end{equation}
where $\varrho_{\rm gy}$ is the gyrocenter charge density \eqref{eq:rho_gy} and the gyrocenter current density is
\begin{equation}
{\bf J}_{\rm gy} \;\equiv\; \sum\;e\;\int\;F\;\frac{d_{\rm gy}{\bf X}}{dt}\;d^{3}p,
\label{eq:J_gy}
\end{equation}
while the gyrocenter parallel momentum vector ${\bf P}_{\|}$ and the gyrocenter parallel-momentum stress tensor $\vb{\Pi}_{\|}$ are
\begin{equation}
\left( {\bf P}_{\|},\; \vb{\Pi}_{{\rm gy}\|} \right) \;=\; \sum\;\int\; F \;\left( \bhat,\; \frac{d_{\rm gy}{\bf X}}{dt}\;\bhat\right)\;
p_{\|}\;d^{3}p.
\label{eq:P_Pi_par}
\end{equation}
Secondly, we introduce the decomposition
\[ -\;\frac{e}{c}\;\nabla{\bf A}^{*}\bdot\frac{d_{\rm gy}{\bf X}}{dt} \;=\; -\;\frac{e}{c}\;\nabla{\bf A}\bdot\frac{d_{\rm gy}{\bf X}}{dt} \;-\;
p_{\|}\;\nabla\bhat\bdot \frac{d_{\rm gy}{\bf X}}{dt} \]
and the Maxwell-tensor divergence
\[ \nabla\bdot\vb{\Pi}_{\rm E} \;=\; -\,\epsilon\;\sum\;e\;\int\; F\;\left\langle{\sf T}_{\rm gy}^{-1}{\bf E}_{1{\rm gc}}\right\rangle\;d^{3}p \;\equiv\; \sum\;e\;\int\;F\;\left( \nabla\Phi_{\rm gy} \;-\; \nabla^{\prime}\Phi_{\rm gy} \right)\;d^{3}p, \]
which combines Eqs.~\eqref{eq:div_Pi_E} and \eqref{eq:Tgy_E}. Thirdly, we use the gyrocenter charge conservation law
\begin{eqnarray}
0 & = & \pd{\varrho_{\rm gy}}{t} \;+\; \nabla\bdot{\bf J}_{\rm gy} \;=\; \pd{}{t}\left(\varrho_{\rm phys} \;+\frac{}{} \nabla\bdot\vb{\cal P}\right)
\;+\; \nabla\bdot\left( {\bf J}_{\rm phys} \;-\; \pd{\vb{\cal P}}{t} \;-\; c\,\nabla\btimes{\bf M} \right) \nonumber \\
 & \equiv & \pd{\varrho_{\rm phys}}{t} \;+\; \nabla\bdot{\bf J}_{\rm phys}
\label{eq:gyro_charge}
\end{eqnarray}
where the physical charge density $\varrho_{\rm phys} \equiv \varrho_{\rm gy} - \nabla\bdot\vb{\cal P}$ is expressed in terms of the gyrocenter charge density \eqref{eq:rho_gy} and the gyrokinetic polarization charge density $-\,\nabla\bdot\vb{\cal P}$, while the physical current density ${\bf J}_{\rm phys} \equiv {\bf J}_{\rm gy} + \partial{\bf P}/\partial t + c\,\nabla\btimes{\bf M}$ is expressed in terms of the gyrocenter current density ${\bf J}_{\rm gy}$, the polarization current density $\partial{\bf P}/\partial t$, and the divergenceless magnetization current density $c\,\nabla\btimes{\bf M}$ (note that the gyrokinetic magnetization ${\bf M}$ is not needed in the present gyrokinetic Vlasov-Poisson theory). Hence, we obtain
\[ \pd{\varrho_{\rm gy}}{t}\;{\bf A} \;+\; \nabla\bdot\left({\bf J}_{\rm gy}\frac{}{}{\bf A} \right) \;=\; {\bf J}_{\rm gy}\bdot\nabla{\bf A} \;=\; 
\sum\;e\;\int\;F\; \frac{d_{\rm gy}{\bf X}}{dt}\bdot\nabla{\bf A}. \]
By inserting these expressions, and using Eq.~\eqref{eq:pbhat_dot}, we obtain the vector gyrokinetic parallel-momentum equation
\begin{equation}
\pd{{\bf P}_{\|}}{t} \;+\; \nabla\bdot\vb{\Pi}_{{\rm gy}\|} \;=\; \sum\;\int\; F\;\frac{d_{\rm gy}(p_{\|}\bhat)}{dt}\;d^{3}p,
\label{eq:Ppar_vector}
\end{equation}
which is clearly expressed as a gyrokinetic Vlasov-moment equation \eqref{eq:chi_moment} with $\chi = p_{\|}\,\bhat$.

Lastly, we take the dot product of Eq.~\eqref{eq:Ppar_vector} with the parallel unit vector $\bhat$ and we obtain the gyrokinetic parallel momentum equation
\begin{equation}
\pd{P_{\|}}{t} \;+\; \nabla\bdot{\bf R}_{\|} \;=\; -\;\sum\;\int\; F \left( \bhat\bdot\nabla H_{\rm gy} \;-\; p_{\|}\;\left(\bhat\bdot\nabla\bhat\right)
\bdot \frac{d_{\rm gy}{\bf X}}{dt} \right) \;d^{3}p \;\equiv\; \sum\;\int\; F \;\frac{d_{\rm gy}p_{\|}}{dt} \;d^{3}p,
\label{eq:Ppar_eq}
\end{equation}
where $(P_{\|}, {\bf R}_{\|}) \equiv ({\bf P}_{\|}\bdot\bhat, \vb{\Pi}_{{\rm gy}\|}\bdot\bhat)$ and we used $(\nabla\bdot\vb{\Pi}_{{\rm gy}\|})\bdot\bhat = \nabla\bdot{\bf R}_{\|}$ with $\nabla\bhat\;:\;\vb{\Pi}_{{\rm gy}\|}^{\top} \equiv 0$ (since $\nabla\bhat\bdot\bhat \equiv 0$). We immediately note that the gyrokinetic parallel momentum equation \eqref{eq:Ppar_eq} has the form of a gyrokinetic Vlasov-moment equation \eqref{eq:chi_moment} with $\chi = p_{\|}$.

\subsection{\label{subsec:gyro_ang}Gyrokinetic Toroidal Angular Momentum}

We are now ready to use the Noether method to derive the gyrokinetic toroidal angular-momentum conservation law associated with the toroidal rotational symmetry of the background magnetic field. First, we insert a virtual infinitesimal toroidal rotation (about the $z$-axis) generated by the displacement 
\begin{equation}
\delta{\bf x} \;\equiv\; \pd{\bf X}{\varphi}\;\delta\varphi \;=\; \delta\varphi\,\wh{\sf z}\btimes{\bf X}
\label{eq:deltax_phi}
\end{equation}
into the primitive gyrokinetic Noether momentum equation \eqref{eq:gyro_mom_prim}. The gyrokinetic angular-momentum conservation law is, therefore, expressed as
\begin{equation}
\pd{P_{\varphi}}{t} \;+\; \nabla\bdot\vb{\Pi}_{\varphi} \;=\; 0,
\label{eq:gyro_angmom_final}
\end{equation}
where we used the axisymmetry condition $\partial^{\prime}{\cal L}_{\rm gy}/\partial\varphi \equiv 0$. Here, the gyrokinetic angular-momentum density is
\begin{equation}
P_{\varphi} \;\equiv\; \sum\;\int\; F\,p_{{\rm gy}\varphi}\;d^{3}p,
\label{eq:P_gyro_phi}
\end{equation}
where the gyrocenter toroidal canonical momentum $p_{{\rm gy}\varphi}$ in defined in Eq.~\eqref{eq:pgy_phi}, and the {\it canonical} gyrokinetic angular-momentum flux
\begin{eqnarray}
\vb{\Pi}_{\varphi} & \equiv & \vb{\Pi}_{\rm E}\bdot\pd{\bf X}{\varphi} \;+\; {\bf R}_{\varphi}
\label{eq:T_gyro_phi}
\end{eqnarray}
is decomposed in terms of the toroidal component of the Maxwell stress tensor \eqref{eq:Pi_E_def} and the toroidal component of the gyrocenter Reynolds stress tensor
\begin{equation}
{\bf R}_{\varphi} \;\equiv\; \sum\int\,F\;\frac{d_{\rm gy}{\bf X}}{dt}\,p_{{\rm gy}\varphi}\;d^{3}p,
\label{eq:Q_phi_def}
\end{equation}
where the gyrocenter velocity is defined in Eq.~\eqref{eq:X_dot}. We note that the connection between the gyrokinetic angular-momentum conservation law \eqref{eq:gyro_angmom_final} and the gyrokinetic Noether momentum equation \eqref{eq:gyro_mom_final} involves the dot product of the gyrokinetic Noether momentum equation \eqref{eq:gyro_mom_final} with $\partial{\bf X}/\partial\varphi$:
\begin{eqnarray}
\pd{{\bf P}}{t}\bdot\pd{\bf X}{\varphi} \;+\; \left(\nabla\bdot\vb{\Pi}\right)\bdot\pd{\bf X}{\varphi} & = & \pd{P_{\varphi}}{t} \;+\;
\nabla\bdot\vb{\Pi}_{\varphi} \;-\; \nabla\left(\pd{\bf X}{\varphi}\right)\;:\; \vb{\Pi}^{\top} \nonumber \\
 & = & \sum\int\;F\;\left( \frac{e}{c}\,\pd{{\bf A}^{*}}{\varphi}\bdot\frac{d_{\rm gy}{\bf X}}{dt} \right) \;d^{3}p,
\label{eq:angmom_varphi}
\end{eqnarray}
where $\vb{\Pi}^{\top}$ denotes the transpose of the canonical gyrokinetic momentum-stress tensor \eqref{eq:T_gyro}. Noting that the dyadic tensor
\eqref{eq:dyadic} is antisymmetric, only the asymmetric part of the canonical gyrokinetic momentum-stress tensor \eqref{eq:T_gyro} contributes to the term
\begin{equation}
\nabla\left(\pd{\bf X}{\varphi}\right)\;:\; \vb{\Pi}^{\top} \;\equiv\; \nabla\left(\pd{\bf X}{\varphi}\right)\;:\; \vb{\Pi}_{\rm gy}^{\top} \;=\; \sum\;\int\;F\;\left[ \frac{e}{c}\;\frac{d_{\rm gy}{\bf X}}{dt}\bdot\nabla\left(\pd{\bf X}{\varphi}\right)\bdot{\bf A}^{*} \right]\; d^{3}p,
\label{eq:Pi_top}
\end{equation}
since the Maxwell stress tensor $\vb{\Pi}_{\rm E}$ is symmetric and $\nabla(\partial{\bf X}/\partial\varphi):\vb{\Pi}_{\rm E}^{\top} \equiv 0$. By substituting Eq.~\eqref{eq:Pi_top} into Eq.~\eqref{eq:angmom_varphi}, we recover the gyrokinetic toroidal angular-momentum conservation law 
\eqref{eq:gyro_angmom_final}:
\begin{equation}
0 \;=\; \pd{P_{\varphi}}{t} \;+\; \nabla\bdot\vb{\Pi}_{\varphi} \;\equiv\; \sum\;\int\;F\;\left\{ \frac{e}{c}\;\frac{d_{\rm gy}{\bf X}}{dt}\bdot
\left[ \pd{{\bf A}^{*}}{\varphi} \;+\; \nabla\left(\pd{\bf X}{\varphi}\right)\bdot{\bf A}^{*} \right] \right\}\;d^{3}p,
\label{eq:Noether_varphi_id}
\end{equation}
where the gyrocenter canonical identity \eqref{eq:varphi_id} reappears.

Lastly, we note that the toroidal component of the Maxwell-tensor divergence \eqref{eq:div_Pi_E} yields
\begin{equation}
\left(\nabla\bdot\vb{\Pi}_{\rm E}\right)\bdot\pd{\bf X}{\varphi} \;=\; \sum\;e\;\int\;F\;\pd{\Phi_{\rm gy}}{\varphi}\;d^{3}p,
\label{eq:div_3}
\end{equation}
where
\begin{equation}
\pd{\Phi_{\rm gy}}{\varphi} \;=\; \epsilon\, \left\langle \pd{\phi_{1{\rm gc}}}{\varphi} \;-\; \epsilon
\left\{ S_{1},\; \pd{\phi_{1{\rm gc}}}{\varphi} \right\}_{\rm gc} \;+\; \cdots \right\rangle \;\equiv\; \epsilon\;\left\langle{\sf T}_{\rm gy}^{-1}\left(\pd{\phi_{1{\rm gc}}}{\varphi}\right)\right\rangle.
\label{eq:Hgy_phi}
\end{equation}
When we combine these contributions into the gyrokinetic angular-momentum conservation law \eqref{eq:gyro_angmom_final}, we finally obtain
\begin{equation}
\pd{P_{\varphi}}{t} \;=\; -\;\nabla\bdot{\bf R}_{\varphi} \;-\; \sum\;\int\,F\;\left( \pd{H_{\rm gy}}{\varphi} \right) d^{3}p,
\label{eq:bigP_phi_dot}
\end{equation}
where $\partial H_{\rm gy}/\partial \equiv e\;\partial\Phi_{\rm gy}/\partial\varphi$. Note that Eq.~\eqref{eq:bigP_phi_dot} can be obtained from the gyrokinetic Vlasov-moment equation \eqref{eq:chi_moment} with $\chi = p_{{\rm gy}\varphi}$: 
\begin{equation}
0 \;=\; \sum\,\int F \left( \frac{d_{\rm gy}p_{{\rm gy}\varphi}}{dt} \;+\; \pd{H_{\rm gy}}{\varphi} \right)\,d^{3}p \;\equiv\; 
\pd{P_{\varphi}}{t} \;+\; \nabla\bdot{\bf R}_{\varphi} \;+\; \sum\;\int\,F\;\left( \pd{H_{\rm gy}}{\varphi} \right) d^{3}p,
\label{eq:Vlasov_moment_varphi}
\end{equation}
as was recently derived by Scott and Smirnov \cite{Scott_Smirnov}. Equation \eqref{eq:bigP_phi_dot} is the general form of the gyrokinetic angular-momentum conservation law and applies to the electromagnetic case as well as the electrostatic case considered here. In the electromagnetic case (to be considered in future work), Eq.~\eqref{eq:bigP_phi_dot} applies to the Hamiltonian formulation of electromagnetic gyrokinetic theory \cite{HLB_1988,Brizard_Hahm}, where all perturbation fields appear in the gyrocenter Hamiltonian and the gyrocenter canonical momentum is still given by Eq.~\eqref{eq:pgy_can}. In the next Section, we will investigate this conservation law in axisymmetric tokamak geometry \eqref{eq:B_tok} for the case of the gyrokinetic Vlasov-Poisson equations.

\section{\label{sec:apply}Gyrokinetic Momentum Conservation in Axisymmetric Tokamak Geometry}

In this Section, we derive the gyrokinetic toroidal angular-momentum conservation law \eqref{eq:bigP_phi_dot} in axisymmetric tokamak geometry \eqref{eq:B_tok}. For this purpose, we introduce the magnetic-surface average
\begin{equation}
\llbracket\;\cdots\;\rrbracket \;\equiv\; \frac{1}{{\cal V}}\;\oint\;(\cdots)\;{\cal J}\;d\vartheta\,d\varphi,
\label{eq:flux_av}
\end{equation}
where ${\cal V}(\psi) \equiv \oint\;{\cal J}\;d\vartheta\,d\varphi$ is the surface integral of the magnetic-coordinate Jacobian. The flux-surface average \eqref{eq:flux_av} satisfies the property
\begin{equation}
\llbracket\nabla\bdot{\bf C}\rrbracket \;\equiv\; \frac{1}{{\cal V}}\;\pd{}{\psi}\left( {\cal V}\;\left\llbracket\frac{}{} {\bf C}\bdot\nabla\psi\right\rrbracket
\right)
\label{eq:mag_surf_def}
\end{equation}
for any vector field ${\bf C}$. In a time-independent axisymmetric tokamak geometry, we note that $\partial/\partial t$ also commutes with magnetic-surface averaging.  

\subsection{\label{subsec:par_tor}Gyrokinetic parallel-toroidal momentum transport equation}

We now present a derivation of the magnetic-surface-averaged gyrokinetic parallel-toroidal transport equation \eqref{eq:bigP_phi_dot} in axisymmetric tokamak geometry, carried out in four steps. 

\subsubsection{Parallel-toroidal canonical momentum}

As a first step, we begin with the gyrocenter toroidal canonical momentum \eqref{eq:pgy_phi}, where we quickly note here that only the parallel component of the gyrocenter toroidal angular velocity $d\varphi/dt$ appears (i.e., the magnetic-drift velocity is absent from $p_{{\rm gy}\varphi}$). We therefore write the gyrokinetic angular-momentum density \eqref{eq:P_gyro_phi} as
\begin{equation}
P_{\varphi} \;=\; \sum\;\int\;F\;p_{{\rm gy}\varphi}\;d^{3}p \;\equiv\; -\;\frac{\psi}{c}\;\varrho_{\rm gy} \;+\; P_{\|\varphi},
\label{eq:Pi_phi}
\end{equation}
where $\varrho_{\rm gy}$ denotes the gyrocenter charge density \eqref{eq:rho_gy} and the parallel-toroidal gyrocenter momentum density is
\begin{equation}
P_{\|\varphi} \;\equiv\; \left( \sum\;\int\,F\;p_{\|}\,d^{3}p\right)\,b_{\varphi}. 
\label{eq:P_par_phi}
\end{equation}

The partial time derivative of Eq.~\eqref{eq:Pi_phi} yields
\begin{eqnarray}
\pd{P_{\varphi}}{t} & = & -\;\frac{\psi}{c}\;\pd{\varrho_{\rm gy}}{t} \;+\; \pd{P_{\|\varphi}}{t},
\label{eq:Piphi_dot_split} 
\end{eqnarray}
so that Eq.~\eqref{eq:bigP_phi_dot} becomes 
\begin{equation}
\pd{P_{\|\varphi}}{t} \;+\; \nabla\bdot{\bf R}_{\varphi} \;=\; \frac{\psi}{c}\;\pd{\varrho_{\rm gy}}{t} \;-\; \sum\;\int\;F\;\pd{H_{\rm gy}}{\varphi}\;d^{3}p,
\label{eq:Piphi_dot}
\end{equation}
which represents the evolution equation for the gyrocenter parallel-toroidal momentum density $P_{\|\varphi}$, where ${\bf R}_{\varphi}$ is defined in 
Eq.~\eqref{eq:Q_phi_def}.

\subsubsection{Magnetic-surface average}

As a second step, we perform the flux-surface average of Eq.~\eqref{eq:Piphi_dot}, which yields 
\begin{eqnarray}
\pd{\llbracket P_{\|\varphi}\rrbracket}{t} & = & -\;\frac{1}{{\cal V}}\;\pd{}{\psi}\left( {\cal V}\frac{}{}\left\llbracket R_{\varphi}^{\psi}\right\rrbracket\right) \;+\; \frac{\psi}{c}\;\pd{\llbracket\varrho_{\rm gy}\rrbracket}{t} \;-\; \sum\;\left\llbracket\int\;F\;\pd{H_{\rm gy}}{\varphi}\;d^{3}p \right\rrbracket,
\label{eq:Pi_phi_av_dot}
\end{eqnarray}
where the surface-averaged radial flux density $R_{\varphi}^{\psi} \equiv {\bf R}_{\varphi}\bdot\nabla\psi$ of gyrocenter toroidal canonical momentum
\begin{equation}
\llbracket R_{\varphi}^{\psi}\rrbracket \;=\; \sum\,\left\llbracket\int\;F\;\frac{d_{\rm gy}\psi}{dt}\;p_{{\rm gy}\varphi}\;d^{3}p\right\rrbracket \;\equiv\; \llbracket R_{\|\varphi}^{\psi}\rrbracket \;-\; \frac{\psi}{c}\;\llbracket J_{\rm gy}^{\psi}\rrbracket,
\label{eq:Gamma_phi}
\end{equation}
is decomposed in terms of the radial component $J_{\rm gy}^{\psi} \equiv {\bf J}_{\rm gy}\bdot\nabla\psi$ of the gyrocenter current density \eqref{eq:J_gy} and the radial flux of gyrocenter parallel-toroidal momentum
\begin{equation}
\llbracket R_{\|\varphi}^{\psi}\rrbracket \;\equiv\; \sum\,\left\llbracket\left(\int\;F\;\frac{d_{\rm gy}\psi}{dt}\;p_{\|}\;d^{3}p\right)\;
b_{\varphi}\right\rrbracket.
\label{eq:Q_parphi}
\end{equation}
By rearranging terms in Eq.~\eqref{eq:Pi_phi_av_dot}, we therefore obtain
\begin{eqnarray}
\pd{\llbracket P_{\|\varphi}\rrbracket}{t} \;+\; \frac{1}{{\cal V}}\;\pd{}{\psi}\left( {\cal V}\;\left\llbracket R_{\|\varphi}^{\psi}\right\rrbracket\right) & = & \sum\;e\;\left\llbracket\int\;F\;\left( \frac{1}{c}\,\frac{d_{\rm gy}\psi}{dt} \;-\; \pd{\Phi_{\rm gy}}{\varphi}\right)\;
d^{3}p \right\rrbracket \nonumber \\
 & = & \sum\,\left\llbracket\int F \;\left[ \frac{d_{\rm gy}}{dt}\,(p_{\|}\,b_{\varphi})\right]\;d^{3}p \right\rrbracket,
\label{eq:Pi_par_phi_dot}
\end{eqnarray}
where we used the surface-averaged gyrocenter charge conservation law \eqref{eq:gyro_charge}:
\begin{equation}
\pd{\llbracket\varrho_{\rm gy}\rrbracket}{t} \;=\; -\;\llbracket\nabla\bdot{\bf J}_{\rm gy}\rrbracket \;\equiv\; -\;\frac{1}{\cal V}\,\pd{}{\psi}\left({\cal V}\;
\left\llbracket J_{\rm gy}^{\psi}\right\rrbracket \right).
\label{eq:gyro_charge_av}
\end{equation}
By using Eq.~\eqref{eq:pb_phi_dot}, we see that Eq.~\eqref{eq:Pi_par_phi_dot} is expressed directly as a gyrocenter Vlasov-moment equation 
\eqref{eq:chi_moment} with $\chi = p_{\|}\,b_{\varphi}$. 

\subsubsection{Gyrocenter quasineutrality condition}

As a third step, we make use of gyrocenter quasineutrality relation 
\begin{equation}
\varrho_{\rm gy} \;\equiv\; \nabla\bdot\vb{\cal P}
\label{eq:quasi_gy}
\end{equation}
between the gyrocenter charge density \eqref{eq:rho_gy} and the gyrokinetic polarization \eqref{eq:pol_epsilon}. The partial time derivative of the surface-averaged gyrocenter quasineutrality condition \eqref{eq:quasi_gy} yields
\begin{equation}
\pd{\llbracket\varrho_{\rm gy}\rrbracket}{t} \;\equiv\; \left\llbracket\nabla\bdot\pd{\vb{\cal P}}{t}\right\rrbracket \;=\; \frac{1}{{\cal V}}\;
\pd{}{\psi}\left( {\cal V}\;\pd{\llbracket{\cal P}^{\psi}\rrbracket}{t}\right),
\label{eq:rho_av}
\end{equation}
where ${\cal P}^{\psi} \equiv \vb{\cal P}\bdot\nabla\psi$ denotes the radial part of the gyrokinetic polarization and $\partial{\cal P}^{\psi}/\partial 
t$ defines the radial component of the gyrocenter polarization-drift current. 

By using the surfaced-averaged gyrocenter charge conservation law \eqref{eq:gyro_charge_av}, we readily obtain the identity
\begin{equation}
\llbracket J_{\rm phys}^{\psi}\rrbracket \;=\; \left\llbracket J_{\rm gy}^{\psi}\right\rrbracket \;+\; \pd{\llbracket{\cal P}^{\psi}\rrbracket}{t}   \;\equiv\; 0,
\label{eq:pol_id}
\end{equation}
which implies that the surfaced-averaged radial component of the physical current $\llbracket{\bf J}_{\rm phys}\bdot\nabla\psi\rrbracket \equiv 0$ vanishes, as is required to preserve the ambipolarity condition \cite{PH_1,PC_2009}. Here, we used the fact that the surface-averaged radial component of the magnetization current density $c\,\nabla\btimes{\bf M}$ vanishes since $\nabla\psi\bdot\nabla\btimes{\bf M} \equiv \nabla\bdot({\bf M}\btimes
\nabla\psi)$ and $\llbracket\nabla\bdot({\bf M}\btimes\nabla\psi)\rrbracket \equiv 0$ by Eq.~\eqref{eq:mag_surf_def}.

\subsubsection{Parallel-toroidal momentum transport equation}

As a fourth and final step, we substitute Eqs.~\eqref{eq:gyro_charge_av} and \eqref{eq:pol_id} in Eq.~\eqref{eq:Pi_phi_av_dot}, and we obtain the gyrokinetic transport equation for the surface-averaged gyrocenter parallel-toroidal momentum density
\begin{eqnarray}
\pd{}{t} \left( \llbracket P_{\|\varphi}\rrbracket \;+\; \frac{1}{c}\;\llbracket{\cal P}^{\psi}\rrbracket \right) \;+\; \frac{1}{{\cal V}}\pd{}{\psi}
\left({\cal V}\;\llbracket R_{\|\varphi}^{\psi}\rrbracket \right) & = & -\; \sum\;e\;\left\llbracket\int\;F\;\pd{\Phi_{\rm gy}}{\varphi}\;d^{3}p \right\rrbracket,
\label{eq:Pi_phi_av_dot_80} 
\end{eqnarray}
which corresponds exactly to Eq.~(80) of the recent work by Scott and Smirnov \cite{Scott_Smirnov}. Here, we see that the transport of the surface-averaged gyrocenter parallel-toroidal momentum density $\llbracket P_{\|\varphi}\rrbracket$ is combined with the surface-averaged radial  gyrokinetic polarization $\llbracket{\cal P}^{\psi}\rrbracket$. In fact, using the magnetic-coordinate identity \eqref{eq:tok_id} and the gyrokinetic polarization \eqref{eq:pol_dipole}, we find that the sum of the gyrocenter parallel-toroidal momentum density and the radial component of the gyrokinetic polarization
\begin{equation}
P_{\|\varphi} \;+\; \frac{1}{c}\,{\cal P}^{\psi} \;\equiv\; \sum\; m\;\int\,F\;\left(\pd{{\bf X}}{\varphi}\bdot\frac{d_{\rm gy}^{(1)}{\bf X}}{dt}
\right)\;d^{3}p \;-\; \frac{1}{c}\,\nabla\psi\bdot\left(\nabla\bdot{\sf Q}\right)
\label{eq:total_tor}
\end{equation}
is expressed in terms of the gyrocenter moment of the toroidal component of the gyrocenter momentum (with the effective gyrocenter potential 
$\Phi_{\rm gy}$ replaced with its first-order contribution $\epsilon\,\langle\phi_{1{\rm gc}}\rangle$) as well as quadrupole contributions to the gyrokinetic polarization. The combination of the gyrocenter canonical toroidal momentum equation \eqref{eq:pgy_phi_dot} and the gyrocenter charge conservation law \eqref{eq:rho_av}, therefore, allows us to recover the gyrocenter Vlasov moment equation for the (truncated) toroidal momentum 
\begin{equation}
m\,\frac{d_{\rm gy}^{(1)}{\bf X}}{dt}\bdot\pd{{\bf X}}{\varphi} \;\equiv\; m\,{\cal R}^{2}\;\frac{d_{\rm gy}^{(1)}\varphi}{dt}.
\label{eq:tor_gy_dot}
\end{equation}
We note here that previous works (e.g., Ref.~\cite{Scott_Smirnov}) have neglected the guiding-center polarization contribution in 
Eq.~\eqref{eq:pol_dipole} and, hence, the first term on the right side of Eq.~\eqref{eq:total_tor} only combines the toroidal components of the gyrocenter parallel vector momentum and the perturbed $E\times B$ velocity and ignores the toroidal component of the magnetic drift velocity.

\subsection{Gyrokinetic parallel-toroidal momentum conservation law}

In their recent paper \cite{Scott_Smirnov}, Scott and Smirnov considered a simplified gyrocenter Hamiltonian obtained in the long-wavelength (zero-Larmor-radius) limit, where the gyrocenter potential \eqref{eq:Psi_gy_def} became $\Phi_{\rm gy} \simeq \epsilon\,\phi_{1} + \frac{1}{2}\,
\epsilon^{2}\,\vb{\rho}_{1\bot}\bdot\nabla_{\bot}\phi_{1}$ and the first-order gyrocenter displacement \eqref{eq:rho1_gy} was
$\langle\vb{\rho}_{1{\rm gy}}\rangle \simeq \vb{\rho}_{1\bot} \equiv -\,(c/B\Omega)\,\nabla_{\bot}\phi_{1}$. In this limit, Scott and Smirnov 
\cite{Scott_Smirnov} found
\[ \pd{\Phi_{\rm gy}}{\varphi} \;\simeq\; \epsilon\;\pd{\phi_{1}}{\varphi} \;-\; \epsilon^{2}\,\vb{\rho}_{1\bot}\bdot\nabla_{\bot}\pd{\phi_{1}}{\varphi}, \] 
and then showed that the last term on the right of Eq.~\eqref{eq:Pi_phi_av_dot_80} can be expressed as a spatial divergence [as can already be seen from Eq.~\eqref{eq:div_3}].

In the present paper, we perform a Taylor expansion of Eq.~\eqref{eq:Hgy_phi} in powers of the gyrocenter displacement $\vb{\rho}_{\epsilon}$, defined in Eq.~\eqref{eq:rho_epsilon}, which yields
\begin{equation}
\pd{H_{\rm gy}}{\varphi} \;\equiv\; \epsilon\,e\;\left\langle {\sf T}_{\epsilon}^{-1}\left(\pd{\phi_{1}}{\varphi}\right)\right\rangle \;=\; \epsilon\,e\;\left(\pd{\phi_{1}}{\varphi} \;+\; \langle\vb{\rho}_{\epsilon}\rangle\bdot\nabla\pd{\phi_{1}}{\varphi} \;+\; \frac{1}{2}\,\langle\vb{\rho}_{\epsilon}
\vb{\rho}_{\epsilon}\rangle\,:\,\nabla\nabla\pd{\phi_{1}}{\varphi} \;+\; \cdots \right),
\label{eq:Ham_gy_varphi}
\end{equation}
where we have combined the guiding-center and gyrocenter phase-space transformations ${\sf T}_{\epsilon}^{-1} \equiv {\sf T}_{\rm gy}^{-1}
{\sf T}_{\rm gc}^{-1}$ [see Eq.~\eqref{eq:delta_epsilon}]. From this expression, we now obtain
\begin{eqnarray}
\sum\;\int\;F\;\pd{H_{\rm gy}}{\varphi}\;d^{3}p & = & \epsilon\,\varrho_{\rm gy}\;\pd{\phi_{1}}{\varphi} \;+\; \epsilon\,\left( \sum\,e\;\int\,F\,
\langle\vb{\rho}_{\epsilon}\rangle\,d^{3}p \right)\bdot\nabla\pd{\phi_{1}}{\varphi} \nonumber \\
 &  &+\; \epsilon\,\left( \sum\,\frac{e}{2}\int\,F\,\langle\vb{\rho}_{\epsilon}\vb{\rho}_{\epsilon}\rangle\,d^{3}p \right) \;:\; \nabla\nabla
\pd{\phi_{1}}{\varphi} \;+\; \cdots
\label{eq:Ham_gy_varphi_sum}
\end{eqnarray}
Next, we substitute the definition of the gyrokinetic polarization \eqref{eq:pol_epsilon} into Eq.~\eqref{eq:Ham_gy_varphi_sum} and, after rearranging terms, we obtain
\begin{eqnarray}
\sum\;\int\;F\;\pd{H_{\rm gy}}{\varphi}\;d^{3}p & = & \epsilon\,\left( \varrho_{\rm gy} \;-\frac{}{} \nabla\bdot\vb{\cal P}\right)\,\pd{\phi_{1}}{\varphi} \;+\;
\nabla\bdot\left[\; \epsilon\,\vb{\cal P}\,\pd{\phi_{1}}{\varphi} \;+\; \epsilon\,\left( \sum\,\frac{e}{2}\int\,F\,\langle\vb{\rho}_{\epsilon}
\vb{\rho}_{\epsilon}\rangle\,d^{3}p \right)\bdot\nabla\pd{\phi_{1}}{\varphi} \;+\; \cdots \;\right] \nonumber \\
 & \equiv & \nabla\bdot\left( \epsilon\,\vb{\cal P}\,\pd{\phi_{1}}{\varphi} \;+\; \epsilon\,{\sf Q}\bdot\nabla\pd{\phi_{1}}{\varphi} \;+\; \cdots
\right),
\label{eq:Ham_gy_varphi_final}
\end{eqnarray}
where we used the gyrocenter quasineutrality condition \eqref{eq:quasi_gy} to show that the last term on the right of Eq.~\eqref{eq:Pi_phi_av_dot_80} can indeed be expressed as an exact spatial divergence involving moments of the gyrocenter displacement $\vb{\rho}_{\epsilon}$. We note that the right side of
Eq.~\eqref{eq:Ham_gy_varphi_final}, which is at least of first order in $\epsilon$, retains all finite-Larmor-radius effects as well as nonlinear corrections, such as the gyroangle-independent part \eqref{eq:rho1_gy} of the first-order gyrocenter displacement $\vb{\rho}_{1{\rm gy}}$, while Scott and Smirnov \cite{Scott_Smirnov} considered only the long-wavelength limit of Eq.~\eqref{eq:Ham_gy_varphi_final}, with the gyrocenter contribution to the gyrokinetic polarization $\vb{\cal P} \simeq -\,\epsilon\,(mnc^{2}/B^{2})\nabla_{\bot}\phi_{1}$ defined in the zero-Larmor-radius limit and all higher-order effects were omitted [see Eqs.~(59) and (91) of Ref.~\cite{Scott_Smirnov}].

When we insert Eq.~\eqref{eq:Ham_gy_varphi_final} into Eq.~\eqref{eq:Pi_phi_av_dot_80}, we obtain the gyrokinetic parallel-toroidal momentum conservation law for the nonlinear gyrokinetic Vlasov-Poisson equations:
\begin{eqnarray}
\pd{}{t}\left( \llbracket P_{\|\varphi}\rrbracket \;+\; \frac{1}{c}\;\llbracket{\cal P}^{\psi}\rrbracket \right) \;+\; \frac{1}{{\cal V}}\pd{}{\psi}\left[{\cal V}\;\left( \llbracket R_{\|\varphi}^{\psi}\rrbracket \;+\; \epsilon\,\left\llbracket {\cal P}^{\psi}\;\pd{\phi_{1}}{\varphi} \;+\; 
{\bf Q}^{\psi}\bdot\nabla\pd{\phi_{1}}{\varphi} \;+\; \cdots \right\rrbracket \right) \;\right] \;=\; 0,
\label{eq:Pi_phi_av_dot_final} 
\end{eqnarray}
where the radial gyrocenter quadrupole vector
\begin{equation}
{\bf Q}^{\psi} \;\equiv\; \nabla\psi\bdot\left( \sum\,\frac{e}{2}\int\,F\,\langle\vb{\rho}_{\epsilon}
\vb{\rho}_{\epsilon}\rangle\,d^{3}p \right) \;=\; {\bf Q}_{0}^{\psi} \;+\; \epsilon\,{\bf Q}_{1{\rm gy}}^{\psi} \;+\; \cdots
\end{equation}
includes guiding-center FLR corrections ${\bf Q}_{0}^{\psi}$ (involving $\vb{\rho}_{0}$) as well as higher-order gyrocenter corrections 
${\bf Q}_{1}^{\psi} + \cdots$ that involve the gyroangle-dependent part of the first-order guiding-center displacement $\vb{\rho}_{{\rm gc}1}$ and the first-order gyrocenter displacement $\vb{\rho}_{{\rm gy}1}$ (e.g., through $\langle\vb{\rho}_{0}\,\vb{\rho}_{1}\rangle$). The gyrokinetic parallel-toroidal momentum conservation law \eqref{eq:Pi_phi_av_dot_final} therefore implies that the transport of gyrocenter parallel-toroidal momentum $\llbracket P_{\|\varphi}\rrbracket$ is intimately connected to the gyrocenter radial polarization $\llbracket {\cal P}^{\psi}\rrbracket$. This connection has recently been investigated through the gyrokinetic Vlasov-moment approach \cite{PHD_2009,McD_2009_PoP,McD_2009_PRL,Scott_Smirnov}. The term 
$\llbracket {\cal P}^{\psi}\;\partial\phi_{1}/\partial\varphi\rrbracket$ in Eq.~\eqref{eq:Pi_phi_av_dot_final} represents the turbulent transport of toroidal momentum \cite{TSH_2007,PHD_2009}, while the term $\llbracket{\bf Q}^{\psi}\bdot\nabla\partial\phi_{1}/\partial\varphi\rrbracket$ involves the radial gradient in the toroidal electric field. 

Lastly, we note that Eq.~\eqref{eq:Pi_phi_av_dot_final} contains the following nonlinear ordering 
\begin{equation}
\epsilon_{\omega}\,\Omega\;\cdot\;(m\,c_{\rm s}{\cal R})\,\ov{P}_{\|} \;\sim\; \frac{1}{a}\;\cdot\;\epsilon\,(k_{\bot}\,\rho_{\rm s})\;e\,\rho_{\rm s}\;
\cdot\; (k_{\bot}\,\rho_{\rm s})\;\frac{T_{\rm e}}{eq},
\label{eq:gyro_Bohm}
\end{equation}
where, on the left side of Eq.~\eqref{eq:gyro_Bohm}, $\Omega^{-1}\,\partial/\partial t \equiv \epsilon_{\omega}$ defines a turbulent time-scale ordering parameter and $\llbracket P_{\|\varphi}\rrbracket \sim (m\,c_{\rm s}{\cal R})\,\ov{P}_{\|}$ defines the normalized parallel momentum, while, on the right side, $|\nabla\psi|\;\partial/\partial\psi \sim 1/a \equiv \epsilon/\rho_{\rm s}$ introduces the ordering parameter $\epsilon \equiv \rho_{\rm s}/a$, 
${\cal P}^{\psi}/|\nabla\psi| \sim \epsilon\,(k_{\bot}\,\rho_{\rm s})\;e\,\rho_{\rm s}$ represents the ordering of the gyrocenter contribution to the radial gyrokinetic polarization \eqref{eq:rho1_gy}, and $\epsilon\,\partial\phi_{1}/\partial\varphi \sim (k_{\bot}\rho_{\rm s}/q)\,T_{\rm e}/e$ represents the ordering of the toroidal derivative of the perturbed electrostatic potential (here, we used $\epsilon\,\phi_{1} \sim \epsilon\,T_{\rm e}
/e$, defined in terms of the electron temperature $T_{\rm e}$, and $\partial/\partial\varphi \sim k_{\bot}a/q \equiv k_{\bot}\rho_{\rm s}/q\epsilon$, where we assume $k_{\bot}\,a \gg k_{\|}\,{\cal R}$). Hence, we obtain from Eq.~\eqref{eq:gyro_Bohm} the following dimensionless ordering
\begin{equation}
\epsilon_{\omega}\;\ov{P}_{\|} \;\sim\; \epsilon^{3}\;(k_{\bot}\,\rho_{\rm s})^{2}\;a/(q{\cal R}).
\label{eq:P_ordering}
\end{equation}
If the turbulent time-scale ordering $\epsilon_{\omega}$ satisfies a gyro-Bohm (gB) ordering \cite{Connor}
\begin{equation}
\epsilon_{\omega} \;\sim\; \frac{1}{\Omega}\;\left.\pd{}{t}\right|_{\rm gB} \;\sim\; \frac{cT_{\rm e}}{e\,B}\; \frac{\epsilon}{a^{2}\Omega} \;\equiv\; \epsilon^{3},
\label{eq:gB_def}
\end{equation}
then the normalized parallel momentum $\ov{P}_{\|}$ satisfies the explicit scaling 
\begin{equation}
\ov{P}_{\|} \;\sim\; (k_{\bot}\,\rho_{\rm s})^{2}\;a/(q{\cal R}), 
\label{eq:ppar_order}
\end{equation}
which involves the fluctuation spectrum in $k_{\bot}$ and physical parameters of the axisymmetric tokamak plasma. For typical modern tokamak plasmas, with $c_{\rm s} \simeq 1000$ km/s (at $T_{\rm e} \simeq 10$ keV) and $a/(q{\cal R}) \simeq 1/10$, we can insert the scaling \eqref{eq:ppar_order} into the experimental upper-limit on intrinsic rotation speeds $U_{\|} \sim \ov{P}_{\|}\,c_{\rm s} < 40$ km/s \cite{intrinsic_speed} and obtain the condition $(k_{\bot}\rho_{\rm s})^{2} < 0.4$ on the fluctuation spectrum, which is consistent with the standard gyrokinetic ordering
$k_{\bot}\rho_{\rm s} \lesssim 1$ \cite{Brizard_Hahm}.

\section{\label{sec:summ}Summary and Discussion}

In the present work, we presented the derivation of the exact gyrokinetic Noether momentum equation \eqref{eq:gyro_mom_final} and the exact gyrokinetic toroidal angular-momentum conservation law \eqref{eq:gyro_angmom_final} associated with the nonlinear gyrokinetic Vlasov-Poisson equations. While previous gyrokinetic momentum transport equations were derived from moments of the gyrokinetic Vlasov equation, the gyrokinetic momentum equations derived here are exact statements which depend on the nonlinear gyrokinetic physics included in the gyrokinetic action functional \eqref{eq:Agyro_def}. Various proofs were also presented throughout the paper to demonstrate the interconnections between the gyrokinetic momentum conservation laws \eqref{eq:varphi_id} and \eqref{eq:Noether_varphi_id}, and several intermediate steps were shown to be expressed as gyrokinetic Vlasov-moment equations \eqref{eq:chi_moment} (see Table \ref{tab:gyro_moment_table}).

\begin{table}
\caption{\label{tab:gyro_moment_table}Gyrokinetic Vlasov-moment Equation \eqref{eq:chi_moment} for various functions $\chi$.} 
\begin{ruledtabular} 
\begin{tabular}{lccccc}
Function $\chi$   & ${\bf p}_{\rm gy}$     & $p_{{\rm gy}\varphi}$           & $p_{\|}\,\bhat$        & $p_{\|}\,b_{\varphi}$ & $p_{\|}$ \\ \hline 
Gyrokinetic Vlasov-moment Equation & Eq.~\eqref{eq:gyro_moment} & Eq.~\eqref{eq:Vlasov_moment_varphi} & Eq.~\eqref{eq:Ppar_vector} & 
Eq.~\eqref{eq:Pi_par_phi_dot} & Eq.~\eqref{eq:Ppar_eq} \\
\end{tabular}
\end{ruledtabular}
\end{table}

The gyrokinetic parallel-toroidal momentum conservation law \eqref{eq:Pi_phi_av_dot_final}, derived in axisymmetric tokamak geometry and averaged over magnetic-flux surfaces, highlights the role of polarization effects that appear explicitly in the surface-averaged gyrokinetic radial polarization 
$\llbracket {\cal P}^{\psi}\rrbracket$ (which includes guiding-center and gyrocenter contributions) and the toroidally-dependent parts that appear in $\llbracket {\cal P}^{\psi}\;\partial\phi_{1}/\partial\varphi + {\bf Q}^{\psi}\bdot\nabla\partial\phi_{1}/\partial\varphi + \cdots\rrbracket$. The major difference between previous works (e.g., Ref.~\cite{Scott_Smirnov}) and our work is that the gyrokinetic polarization \eqref{eq:pol_dipole} considered here includes contributions from the guiding-center and gyrocenter phase-space transformations, while previous works have neglected the guiding-center polarization contribution (which is of the same order as the gyrocenter contribution).

In future work, the exact gyrokinetic momentum and angular momentum conservation laws for the nonlinear gyrokinetic Vlasov-Maxwell equations, where fully electromagnetic fluctuations as well as gyrokinetic polarization and magnetization are taken into account, will be derived by the Noether method. Because the gyrokinetic Noether momentum equation \eqref{eq:gyro_mom_final} was derived in general magnetic geometry, applications of the present work to stellarator plasmas \cite{PH_1,PH_2,PH_3} and other nonaxisymmetric magnetic geometries could also be considered. 

\acknowledgments

One of us (AJB) acknowledges stimulating discussions with J.~A.~Krommes and G.~W.~Hammett during a recent visit to PPPL in May 2011 as well as during 
the workshop ``Gyrokinetics in Laboratory and Astrophysical Plasmas'' in August 2010 at the Isaac Newton Institute (Cambridge, UK), where some of this work was completed.

Work by AJB was supported by a U.~S.~Dept.~of Energy grant under contract No.~DE-FG02-09ER55005, while work by NT was supported by CNRS, ANR EGYPT, and Euratom-CEA (contract EUR 344-88-1 FUA F).

\appendix

\section{\label{sec:Noether}Noether Method}

In this brief Appendix, we use the Noether method to derive the energy-momentum and angular-momentum conservation laws \cite{Hill,Rosen,Goldstein}. In order to simplify the presentation, we assume the Lagrangian density ${\cal L}(\psi^{a}, \partial_{\mu}\psi^{a}; {\bf x}, t)$ depends on a $N$-component field $\psi^{a} = (\psi^{1},...,\psi^{N})$ and its space-time derivatives $\partial_{\mu}\psi^{a} = (c^{-1}\partial_{t}\psi^{a}, \nabla\psi^{a})$ as well as possible space-time dependence due to external fields. Furthermore, we assume that each component $\psi^{a}$, which is to be varied independently of each other, satisfies the Euler-Lagrange equation
\begin{equation}
\pd{}{x^{\mu}} \left( \pd{\cal L}{(\partial_{\mu}\psi^{a})} \right) \;=\; \pd{\cal L}{\psi^{a}}.
\label{eq:EL_a}
\end{equation}

In this case, the Noether method is based on the Noether equation
\begin{eqnarray}
\delta{\cal L} & = & \delta\psi^{a}\;\pd{\cal L}{\psi^{a}} \;+\; \pd{\delta\psi^{a}}{x^{\mu}}\;\pd{\cal L}{(\partial_{\mu}\psi^{a})} \;=\;
\delta\psi^{a}\;\left[ \pd{\cal L}{\psi^{a}} \;-\; \pd{}{x^{\mu}} \left( \pd{\cal L}{(\partial_{\mu}\psi^{a})} \right) \right] \;+\;
\pd{}{x^{\mu}} \left[ \pd{\cal L}{(\partial_{\mu}\psi^{a})}\;\delta\psi^{a} \right] \nonumber \\
 & \equiv & \pd{}{x^{\mu}} \left[ \pd{\cal L}{(\partial_{\mu}\psi^{a})}\;\delta\psi^{a} \right],
\label{eq:Noether_eq}
\end{eqnarray}
where the variations $\delta\psi^{a}$ and $\delta{\cal L}$ will now be expressed in terms of infinitesimal space-time translations (energy-momentum) or infinitesimal rotations (angular-momentum).

\subsection{Energy-momentum Conservation Law}

In order to derive the energy-momentum conservation law through the Noether method, we consider arbitrary infinitesimal space-time translations generated by $\delta x^{\nu}$:
\begin{equation}
x^{\nu} \;\rightarrow\; x^{\nu} \;+\; \delta x^{\nu}.
\label{eq:delta_trans}
\end{equation}
In this case, the variations $\delta\psi^{a}$ and $\delta{\cal L}$ are expressed as
\begin{eqnarray}
\delta\psi^{a} & = & -\;\delta x^{\nu}\;\partial_{\nu}\psi^{a}, \label{eq:psi_trans} \\
\delta{\cal L} & = & -\;\partial_{\nu}\left(\delta x^{\nu}\frac{}{}{\cal L}\right) \;+\; \delta x^{\nu}\;\partial_{\nu}^{\prime}{\cal L},
\label{eq:L_trans}
\end{eqnarray}
where the partial derivative $\partial_{\nu}^{\prime}{\cal L}$ is calculated with the fields $\psi^{a}$ held constant. If we substitute the variations
\eqref{eq:psi_trans}-\eqref{eq:L_trans} into the Noether equation \eqref{eq:Noether_eq}, we obtain the energy-momentum equation \cite{Goldstein}
\begin{equation}
\partial_{\mu}T^{\mu\nu} \;=\; \partial^{\prime\nu}{\cal L},
\label{eq:en_mom_eq}
\end{equation}
where the {\it canonical} energy-momentum tensor $T^{\mu\nu}$ is defined as
\begin{equation}
T^{\mu\nu} \;\equiv\; g^{\mu\nu}\;{\cal L} \;-\; \pd{\cal L}{(\partial_{\mu}\psi^{a})}\;\partial^{\nu}\psi^{a}.
\label{eq:en_mom_tensor}
\end{equation}
The Noether Theorem now states that if the Lagrangian density ${\cal L}$ is invariant under space-time translations along the $x^{\lambda}$ coordinate
(i.e., $\partial^{\prime\lambda}{\cal L} \equiv 0$), then the components $T^{\mu\lambda}$ satisfy the energy-momentum conservation law
\begin{equation}\
0 \;=\; \partial_{\mu}T^{\mu\lambda} \;=\; c^{-1}\;\partial_{t}T^{0\lambda} \;+\; \partial_{j}T^{j\lambda}.
\label{eq:en_mom_cons}
\end{equation}
For time-translation invariance $(\lambda = 0)$, we define the energy density ${\cal E} \equiv T^{00}$ and the energy-density flux $S^{j} \equiv c\,
T^{j0}$, and Eq.~\eqref{eq:en_mom_cons} yields the energy conservation law
\begin{equation}
\pd{\cal E}{t} \;+\; \nabla\bdot{\bf S} \;=\; 0.
\label{eq:en_cons}
\end{equation}
For space-translation invariance $(\lambda = j)$, we define the momentum density $P^{j} \equiv T^{0j}/c$ and the stress tensor $T^{ij} \equiv \Pi^{ij}$, and Eq.~\eqref{eq:en_mom_cons} yields the momentum conservation law
\begin{equation}
\pd{{\bf P}}{t} \;+\; \nabla\bdot\vb{\Pi} \;=\; 0.
\label{eq:mom_cons}
\end{equation}

\subsection{Angular-momentum Conservation Law}

In order to derive the angular-momentum conservation law through the Noether method, we consider arbitrary infinitesimal rotations generated by the antisymmetric tensor $\delta \Lambda^{\nu\sigma} \equiv -\,\delta \Lambda^{\sigma\nu}$:
\begin{equation}
x^{\nu} \;\rightarrow\; x^{\nu} \;+\; \delta\Lambda^{\nu\sigma}\,x_{\sigma}.
\label{eq:delta_rot}
\end{equation}
In this case, the variations $\delta\psi^{a}$ and $\delta{\cal L}$ are expressed as
\begin{eqnarray}
\delta\psi^{a} & = & -\; \frac{1}{2}\,\delta\Lambda^{\nu\sigma}\;\left(x_{\sigma}\;\partial_{\nu}\psi^{a} \;-\frac{}{} x_{\nu}\;\partial_{\sigma}
\psi^{a} \right) \;+\; \Delta\psi^{a}, \label{eq:psi_rot} \\
\delta{\cal L} & = & -\;\partial_{\nu}\left(\delta\Lambda^{\nu\sigma}\,x_{\sigma}\frac{}{}{\cal L}\right) \;+\; \frac{1}{2}\,\delta\Lambda^{\nu\sigma}\;\left(x_{\sigma}\;\partial^{\prime}_{\nu}{\cal L} \;-\frac{}{} x_{\nu}\;
\partial^{\prime}_{\sigma}{\cal L}\right).
\label{eq:L_rot}
\end{eqnarray}
In Eq.~\eqref{eq:psi_rot}, the Lagrangian variation $\Delta\psi^{a}$ associated with the infinitesimal rotation \eqref{eq:delta_rot} is defined as
\begin{equation}
\Delta\psi^{a} \;\equiv\; \frac{1}{2}\;\delta\Lambda^{\mu\nu}\;\omega_{\mu\nu}^{a},
\label{eq:psi_Lag}
\end{equation}
where the antisymmetric tensor $\omega_{\mu\nu}^{a}$ (whose role will be clarified below) depends on the field components $(\psi^{1},...,\psi^{N})$. 

If we substitute the variations \eqref{eq:psi_rot}-\eqref{eq:L_rot} into the Noether equation \eqref{eq:Noether_eq}, we obtain the angular-momentum equation
\begin{equation}
\partial_{\mu}J^{\mu[\nu\sigma]} \;=\; x^{\sigma}\,\partial^{\prime\nu}{\cal L} \;-\; x^{\nu}\,\partial^{\prime\sigma}{\cal L},
\label{eq:ang_mom_eq}
\end{equation}
where the {\it total} angular-momentum tensor $J^{\mu[\nu\sigma]} \equiv -\,J^{\mu[\sigma\nu]}$ is defined as
\begin{equation}
J^{\mu[\nu\sigma]} \;\equiv\; L^{\mu[\nu\sigma]} \;+\; S^{\mu[\nu\sigma]}.
\label{eq:total_ang_mom}
\end{equation}
Here, the {\it orbital} angular-momentum tensor
\begin{equation}
L^{\mu[\nu\sigma]} \;\equiv\; T^{\mu\nu}\;x^{\sigma} \;-\; T^{\mu\sigma}\;x^{\nu},
\label{eq:orb_ang_mom}
\end{equation}
where the canonical energy-momentum tensor $T^{\mu\nu}$ is defined in Eq.~\eqref{eq:en_mom_tensor}, satisfies the equation 
\begin{eqnarray}
\partial_{\mu}L^{\mu[\nu\sigma]} & = & \left( x^{\sigma}\,\partial^{\prime\nu}{\cal L} \;-\frac{}{} x^{\nu}\,\partial^{\prime\sigma}{\cal L} \right)
\;+\; \left( T^{\sigma\nu} \;-\; T^{\nu\sigma} \right),
\label{eq:orb_ang_eq}
\end{eqnarray}
The {\it spin} angular-momentum tensor
\begin{equation}
S^{\mu[\nu\sigma]} \;\equiv\; \pd{\cal L}{(\partial_{\mu}\psi^{a})}\;\omega^{a[\nu\sigma]}
\label{eq:spin_ang_mom}
\end{equation}
satisfies the equation
\begin{equation}
\partial_{\mu}S^{\mu[\nu\sigma]} \;\equiv\; T^{\nu\sigma} \;-\; T^{\sigma\nu},
\label{eq:spin_antisymmetry}
\end{equation}
which explicitly links the antisymmetric part of the energy-momentum tensor \eqref{eq:en_mom_tensor} with the antisymmetric tensor 
$\omega^{a[\nu\sigma]}$. When energy-momentum is conserved (i.e., $\partial_{\mu}T^{\mu\nu} \equiv 0$ for $\nu = 0,...,3$), the total angular-momentum \eqref{eq:total_ang_mom} is also conserved, while the orbital and spin angular momenta satisfy the equation
\begin{eqnarray}
\partial_{\mu}L^{\mu[\alpha\beta]} & = & T^{\beta\alpha} \;-\; T^{\alpha\beta} \;\equiv\; -\;\partial_{\mu}S^{\mu[\alpha\beta]}.
\label{eq:orb_spin_eq}
\end{eqnarray}
Hence, the angular-momentum conservation law $\partial_{\mu}J^{\mu[\alpha\beta]} = 0$ relates the orbital angular-momentum density $L^{0[\alpha\beta]}$ with the spin angular-momentum density $S^{0[\alpha\beta]}$ in a form that is similar to Eq.~\eqref{eq:Pi_phi_av_dot_final}.

Lastly, we note that the canonical energy-momentum tensor $T^{\mu\nu}$ can also be symmetrized \cite{Belinfante_1939,Belinfante_1940,Rosenfeld} directly by adding the term $\partial_{\alpha}\Sigma^{[\alpha\mu]\nu}$ (where $\Sigma^{[\alpha\mu]\nu} \equiv -\,\Sigma^{[\mu\alpha]\nu}$) and require that the new energy-momentum tensor be symmetric:
\begin{equation}
\ov{T}^{\mu\nu} \;\equiv\; T^{\mu\nu} \;+\; \partial_{\alpha} \Sigma^{[\alpha\mu]\nu} \;\equiv\; \ov{T}^{\nu\mu}, 
\label{eq:new_T_munu}
\end{equation}
while satisfying the same energy-momentum conservation law $\partial_{\mu}\ov{T}^{\mu\nu} \equiv \partial_{\mu}T^{\mu\nu}$ (since 
$\partial^{2}_{\mu\alpha} \Sigma^{[\alpha\mu]\nu} \equiv 0$). The Belinfante-Rosenfeld solution \cite{McLennan} to Eq.~\eqref{eq:spin_antisymmetry} is expressed as
\begin{equation}
\Sigma^{[\alpha\mu]\nu} \;\equiv\; \frac{1}{2} \left( S^{\nu[\alpha\mu]} \;+\; S^{\mu[\alpha\nu]} \;-\frac{}{} S^{\alpha[\mu\nu]} \right),
\label{eq:BR_sol}
\end{equation}
so that $\Sigma^{[\alpha\mu]\nu} - \Sigma^{[\alpha\nu]\mu} \equiv -\,S^{\alpha[\mu\nu]}$. Hence, the spin angular-momentum density $S^{\alpha[\mu\nu]}$, which is driven by the asymmetry of the canonical energy-momentum tensor [Eq.~\eqref{eq:spin_antisymmetry}], can also be used to obtain a symmetric energy-momentum tensor \eqref{eq:new_T_munu}, where $\Sigma^{[\alpha\mu]\nu}$ is defined by Eq.~\eqref{eq:BR_sol}.

\end{document}